\documentclass[aps,prl,preprint,superscriptaddress,amsmath,amssymb,floatfix]{revtex4-2}
\usepackage{graphicx,bm,xcolor,hyperref,microtype}

\def\Blue#1{\textcolor{blue}{ #1}}

\hypersetup{colorlinks=true,citecolor=blue,linkcolor=blue,urlcolor=blue}
\newcommand{\kk}{\bm{k}}
\newcommand{\rr}{\bm{r}}
\newcommand{\NN}{\bm{N}}
\newcommand{\bA}{\bm{A}}
\newcommand{\PT}{\mathcal{PT}}
\newcommand{\TR}{\mathcal{T}}

\begin{document}
\title{Light-Induced Odd-Parity \textit{h}-Wave Magnetism in Penrose Quasicrystals}
\author{Yiming Li}
\affiliation{School of Physics, Peking University, Beijing 100871, China}
\author{Huaqing Huang}
\email[Corresponding author: ]{huaqing.huang@pku.edu.cn}
\affiliation{School of Physics, Peking University, Beijing 100871, China}
\affiliation{Collaborative Innovation Center of Quantum Matter, Beijing 100084, China}
\affiliation{Center for High Energy Physics, Peking University, Beijing 100871, China}
\date{October 4, 2026}

\begin{abstract}
Noncrystallographic rotational symmetry offers a route to unconventional spin splitting beyond the constraints of periodic crystals. We show that circularly polarized light induces odd-parity \(h\)-wave magnetism in a collinear antiferromagnetic Penrose quasicrystal without spin-orbit coupling. Light-induced chiral hopping, generated by virtual two-step processes, combines with staggered exchange to lift the spin degeneracy of the undriven system. Quasicrystalline spin-group symmetry enforces odd-parity spin splitting and selects \(h\)-wave as its leading angular harmonic. An analytical effective quasi-band theory relates this splitting to the geometry of local hopping paths and predicts its reversal upon switching the light helicity. Momentum-resolved spectral calculations reveal the characteristic alternating spin contrast, for which we propose spin-resolved time- and angle-resolved photoemission spectroscopy as an experimental probe. Our results establish a mechanism for optically controlling higher-harmonic odd-parity magnetism and demonstrate how quasicrystalline order extends Floquet engineering beyond crystallographic symmetry constraints.
\end{abstract}
\maketitle

\Blue{\textit{Introduction.}---}
Unconventional collinear magnetism has emerged as a new class of magnetic order characterized by momentum-dependent spin splitting without spin-orbit coupling (SOC)~\cite{jungwirth2025altermagnetismunconventionalspinorderedphase,  Song2025NatureRevMater, adfm.202409327, luo2026unconventionalmagnetismsymmetryclassification}. A prominent example is altermagnetism, which combines compensated magnetic moments with even-parity spin splitting exhibiting characteristic $d$-, $g$-, or higher-wave angular harmonics~\cite{sciadv.aaz8809,JunweiLiu2021Natcomm, PhysRevX.12.031042, PhysRevX.12.040501, PhysRevX.12.011028, PhysRevLett.131.256703, PhysRevLett.132.036702,  Kremps2024Nature, Reichlova2024Observation, PhysRevX.15.021083}. More recently, odd-parity magnetism has attracted growing attention, with spin splitting satisfying $\Delta_\lambda(-\kk)=-\Delta_\lambda(\kk)$, where $\Delta_\lambda$ denotes the energy difference between opposite-spin states in a matched branch $\lambda$~\cite{hellenes2024pwavemagnets, PhysRevLett.133.236703, yu2025oddparitymagnetismdrivenantiferromagnetic, Song2025p-wave, yamada2025metallic, chakraborty2025highly, CongjunWu2026PRB, ydjj-6r9l,7cpd-mv33,kudo2026oddparitymagnetismquantumgeometry, Yanzhongbo2026odd_parity_orbital_order,jungwirth2024altermagnetsbeyondnodalmagneticallyordered}. Such unconventional spin splitting can exhibit $p$-, $f$-, and higher-wave angular harmonics, providing additional possibilities for spin-dependent electronic responses. Floquet engineering offers a promising route to dynamically control these magnetic states by breaking the symmetries that protect spin degeneracy \cite{Oka2019,Zhan2024}. Recent studies have proposed light-induced $p$- and $f$-wave spin splitting in collinear antiferromagnets~\cite{Ruanjiawei2026_PRL,Wangrui2026_PRL,Li2026FloquetSpin,acs.nanolett.6c00927, wnqs-3djt, Zeng2026SpinGroup, Zeng2026ChernReconstruction, LinVila2026Sublattice}. Nevertheless, realizing higher-harmonic odd-parity magnetism through optical driving remains challenging, motivating the search for systems with unconventional rotational symmetries.

Quasicrystals provide a distinctive platform for exploring unconventional magnetism beyond crystallographic symmetry constraints \cite{janssen2018aperiodic,deloudi2009crystallography, fukamichi1999magnetic, FISHER2000223,
 Vedmedenko_S0217984905009390,Vedmedenko21022006,Hippert01052008, stadnik2013magnetic}. Unlike periodic crystals, they possess long-range quasiperiodic order and noncrystallographic rotational symmetries, such as five-, eight-, and tenfold rotations. Long-range antiferromagnetic order has recently been observed in quasicrystalline materials~\cite{Tamura2021, Tamura2025}, while theoretical studies have predicted ferromagnetic, ferrimagnetic, and altermagnetic phases in quasiperiodic lattices~\cite{godreche1986magnetic, PhysRevLett.90.137203, PhysRevLett.90.177205, PhysRevB.71.115101, PhysRevB.75.212407, jagannathan2009antiferromagnetism, PhysRevB.96.214402, liyiming_2026, src1-b2xt, shao2026classification, ji2026neelorderedmagneticphases}. In particular, quasicrystalline altermagnetism demonstrates that noncrystallographic rotational symmetry can support unconventional higher-harmonic spin splitting. The fivefold rotational symmetry of the Penrose quasicrystal is especially intriguing: when preserved within a spin sector, it excludes the lower odd-parity $p$- and $f$-wave harmonics, making $h$-wave the leading symmetry-allowed angular structure. This raises two fundamental questions: Can optical driving generate such higher-harmonic odd-parity magnetism in a quasicrystal? If so, how can its characteristic spin splitting be identified and experimentally detected in the absence of conventional Bloch bands?

In this Letter, we demonstrate that circularly polarized light (CPL) induces odd-parity $h$-wave spin splitting in a collinear antiferromagnetic Penrose quasicrystal without SOC. The underlying mechanism originates from light-induced chiral hopping generated by virtual two-step processes, which, together with staggered magnetic exchange, lifts the spin degeneracy of the undriven system. Momentum-resolved spectral calculations reveal a characteristic fifth-harmonic spin contrast, whose odd parity and angular structure are established by quasicrystalline spin-group symmetry. We further develop an analytical effective quasi-band theory that connects the spin splitting to the geometry of local hopping paths and predicts its reversal upon switching the light helicity. Finally, we propose spin-resolved time- and angle-resolved photoemission spectroscopy (trARPES) to detect the characteristic spin splitting. Our findings establish a microscopic mechanism for optically controlling higher-harmonic odd-parity magnetism and reveal new opportunities for Floquet engineering enabled by noncrystallographic rotational symmetry.

\begin{figure}[t]
\centering
\includegraphics[width=1\columnwidth]{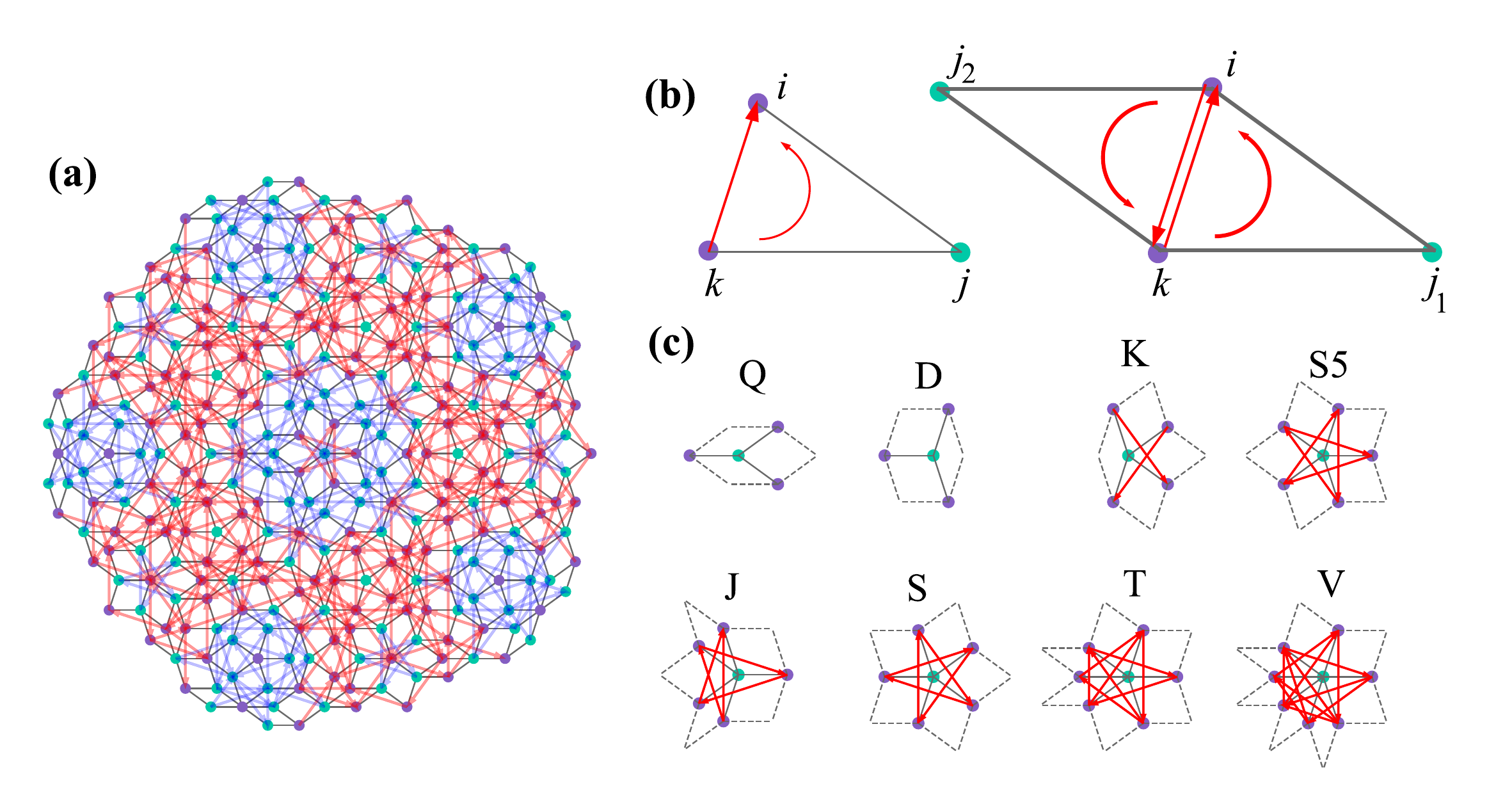}
\caption{\label{fig:mechanism}
{Light-induced chiral virtual hopping in a Penrose quasicrystal.}
(a) Bipartite Penrose tiling, with red and blue arrows indicating effective virtual hoppings within the $A$ and $B$ magnetic sublattices, respectively.
(b) A virtual two-step path $i-j-k$ connects sites on the same sublattice. Contributions through the two intermediate vertices of a rhombus cancel along its diagonal.
(c) Distinct local vertex environments determine the multiplicities of surviving chiral hopping paths~\cite{DEBRUIJN198153, PhysRevB.34.6924, PhysRevB.40.11185}.
}
\end{figure}

\Blue{\emph{Model and light-induced virtual hopping.---}}
{The Penrose tiling can be constructed by projecting selected sites of a five-dimensional (5D) hypercubic lattice onto the 2D physical plane~\cite{DEBRUIJN198153}. Each parent-lattice site is labeled by an integer vector $\bm n_i=\sum_{\nu=1}^{5}n_{i\nu}\bm e_\nu$, where $\bm e_\nu$ denotes the $\nu$-th hypercubic basis vector and $n_{i\nu}\in\mathbb Z$. Its physical-space position is $\bm r_i=S\bm n_i$, where $S$ is the projection matrix (see Supplemental Material (SM) Sec.~I~\cite{SuppMat}).}
The parity $\chi_i=(-1)^{\sum_{\nu=1}^5 n_{i\nu}}$ defines magnetic sublattices $A$ and $B$, connected by edges of length $a$ in 2D physical plane. The acceptance window, a bounded region in perpendicular space, selects the retained parent sites and determines the frequencies of local vertex environments 
(Fig.~\ref{fig:mechanism}; SM Sec.~I~\cite{SuppMat}). In the absence of light, electrons hop along edges with amplitude $-t_1$ and experience a staggered exchange field $\chi_i\NN$, where $\NN=N\hat z$. Without SOC, spin along $z$ is conserved, with $s=\pm 1$ denoting spin up and down. The undriven antiferromagnet serves as the reference state, whose bulk spin degeneracy is protected by the symmetries discussed below.

We apply CPL propagating normal to the plane, with vector potential $\bA(t)=A_0(\cos\omega t,\eta\sin\omega t)$, where $A_0$, $\omega$, and $\eta=\pm1$ denote the amplitude, angular frequency, and helicity, respectively. Incorporating the orbital coupling through the Peierls substitution gives
\begin{equation}
H(t)=N\sum_i\chi_i c_i^\dagger\sigma_zc_i
-t_1\sum_{\langle ij\rangle}
e^{i\frac{e}{\hbar}\bA(t)\cdot(\rr_i-\rr_j)}
c_i^\dagger c_j,
\label{eq:Ht}
\end{equation}
where $c_i=(c_{i\uparrow},c_{i\downarrow})^T$, $\sigma_z$ is a spin Pauli matrix, and $e>0$ is the elementary charge. The bond sum includes both orientations, ensuring Hermiticity. We assume a spatially uniform optical field and a fixed N\'eel background.

In the off-resonant high-frequency regime, where $\hbar\omega$ exceeds the relevant electronic energy scales, the driven system is described by an effective Floquet Hamiltonian. Expanding $H(t)=\sum_n H_n e^{-in\omega t}$, with
$H_n=(\omega/2\pi)\int_0^{2\pi/\omega}dt\,H(t)e^{in\omega t}$,
the leading van Vleck expansion yields~\cite{Eckardt2015}
\begin{equation}
H_{\rm F}=H_0+\sum_{n>0}
\frac{[H_{-n},H_n]}{n\hbar\omega}
+O(\omega^{-2}).
\label{eq:FM}
\end{equation}
The cycle-averaged term $H_0$ renormalizes hopping as $t_1\to t_1J_0(\alpha_0)$, where $\alpha_0=eA_0a/\hbar$ and $J_n$ is the Bessel function of the first kind. For $n\neq0$, the Fourier components of the hopping are proportional to $i^nJ_n(\alpha_0)e^{in\eta\theta_{ij}}$, where $\theta_{ij}$ is the polar angle of $\rr_i-\rr_j$. Details of the Fourier derivation are provided in SM Sec.~II~\cite{SuppMat}.

The commutator in Eq.~\eqref{eq:FM} describes virtual two-step hopping through an intermediate vertex $j$, accompanied by the absorption and emission of $n$ photons. Interference between the two operator orderings generates an imaginary, helicity-dependent hopping. Combining a path $i$-$j$-$k$ with its Hermitian conjugate gives
\begin{align}
H_{\rm F}^{i-j-k}
&=i\eta\Lambda(\delta)
(c_i^\dagger c_k-c_k^\dagger c_i),\nonumber\\
\Lambda(\delta)
&=\frac{2t_1^2}{\hbar\omega}
\sum_{n>0}\frac{J_n^2(\alpha_0)}{n}\sin(n\delta),
\label{eq:H2}
\end{align}
where $\delta=\theta_{kj}-\theta_{ij}$ is the oriented angle between the two bonds directed outward from $j$. Since each elementary hop switches sublattice, the effective hopping connects sites with parallel local magnetic moments. Its imaginary amplitude changes sign upon reversing either the path orientation or the light helicity.

Crucially, contributions from different intermediate vertices interfere coherently. For endpoints across a rhombus diagonal, the two virtual paths yield opposite amplitudes and cancel [Fig.~\ref{fig:mechanism}(b)]. The surviving chiral virtual hoppings therefore depend on the local vertex environments, as illustrated in Fig.~\ref{fig:mechanism}(c), rather than solely on the endpoint pairs. Their enumeration and weights of distinct types of local vertex environments and path counting are detailed in SM Secs.~I and II~\cite{SuppMat}.

In the weak-drive limit, \(\Lambda(\delta) = t_1^2\alpha_0^2\sin\delta/(2\hbar\omega) + O(\alpha_0^4)\). Although the induced hopping is spin-independent, it connects sites within the same magnetic sublattice. Combined with staggered exchange, which couples spin to sublattice, the difference between sublattice-dependent hopping contributions generates momentum-dependent spin splitting. This mechanism is the quasiperiodic counterpart of Haldane-type hopping in crystalline Floquet magnets and related sublattice-current mechanisms~\cite{PhysRevLett.61.2015, Ruanjiawei2026_PRL, LinVila2026Sublattice, ydjj-6r9l, Zeng2026ChernReconstruction, Zeng2026SpinGroup, eto2026oddparitymagnonshaldanehubbardmodel}. Here, the chiral hoppings are generated optically, with their path interference governed by Penrose geometry, without requiring spontaneous current order. In contrast, the cycle-averaged renormalization of real hoppings alone cannot lift the spin degeneracy.

\begin{figure*}[t]
\centering
\includegraphics[width=1\textwidth]{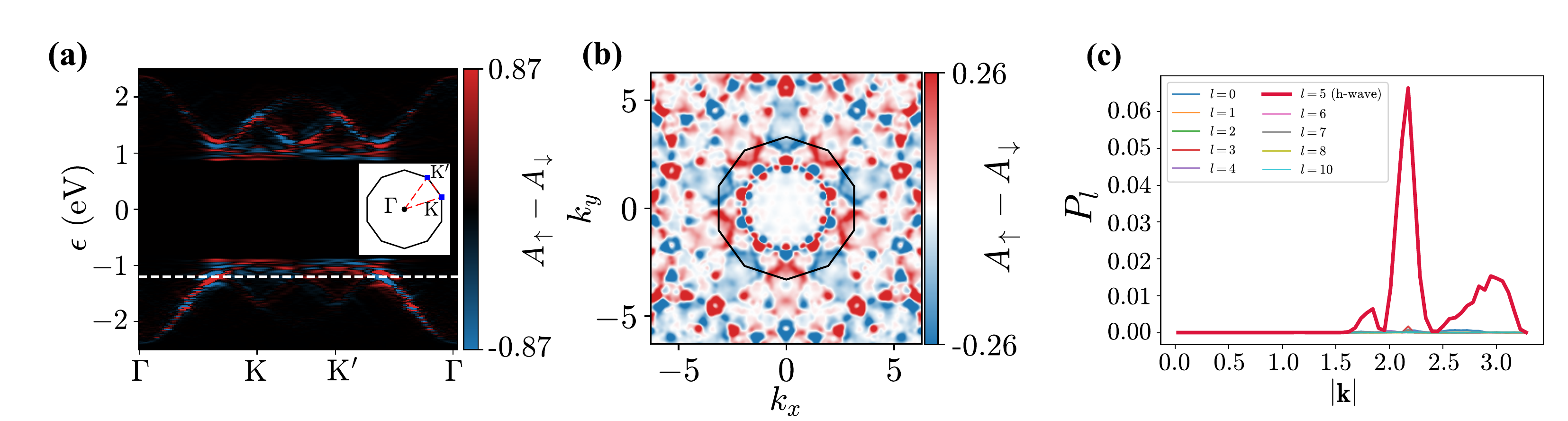}
\caption{\label{fig:spectrum} Spin-resolved spectral functions of CPL-induced odd-parity \(h\)-wave magnetism in the Penrose quasicrystal. (a) Spin-difference spectral cuts along the indicated momentum paths. Inset: decagonal pseudo-Brillouin zone used as a plotting domain, not a primitive reciprocal unit cell. (b) \(\Delta\mathcal A=\mathcal{A}_\uparrow-\mathcal{A}_\downarrow\) in the momentum plane at \(\varepsilon=-1.20\,{\rm eV}\). (c)
Angular weights \(P_\ell\) versus radius \(k\) for the map in (b). The direct spectral calculation uses \(N_{\rm site}=1387\), $t_1=N=1\,{\rm eV}$, $\omega = 20 \,{\rm rad/fs}$, $\alpha_0 = 1.5$, and $\Gamma=0.01\,{\rm eV}$.}
\end{figure*}

\Blue{\emph{Momentum-resolved spectral signature.---}}
We first calculate the spectrum of the finite real-space Hamiltonian \(H_{\rm F}\), fully accounting for local environments and interference between hopping paths. Although momentum is not a conserved quantum number in quasicrystals, projection onto plane-wave states provides a momentum-resolved spectral function without assuming translational symmetry:
\begin{align}
|\kk,s\rangle&=\frac{1}{\sqrt{N_{\rm site}}}
 \sum_j e^{i\kk\cdot\rr_j}|j,s\rangle,\nonumber\\
{\mathcal A}_s(\kk,\varepsilon)&=-\frac1\pi\operatorname{Im}
 \langle\kk,s|(\varepsilon+i\Gamma-H_{\rm F})^{-1}|\kk,s\rangle .
\label{eq:spectral}
\end{align}
Here \(N_{\rm site}\) is the number of vertices, and \(\Gamma>0\) denotes the Lorentzian energy broadening. The resulting spectral function resolves the eigenenergies weighted by their spin-polarized plane-wave overlaps, revealing dispersive features whose intensities encode the diffraction and local-environment structure of the quasicrystal.

Figure~\ref{fig:spectrum}(a) shows dispersive spin-difference spectra along adjacent radial momentum paths from $\Gamma$, which exhibit opposite spin characters. At fixed energy, \(\Delta\mathcal A=\mathcal A_\uparrow-\mathcal A_\downarrow\) displays ten alternating angular sectors in the 2D momentum plane [Fig.~\ref{fig:spectrum}(b)], consistent with odd-parity \(h\)-wave magnetism. This spin contrast vanishes in the undriven bulk limit, confirming its light-induced origin.

To quantify the angular pattern, we write \(\kk=k(\cos\theta,\sin\theta)\), where \(\theta\) is the momentum polar angle, distinct from the bond angles in Eq.~\eqref{eq:H2}, and define
\begin{equation}
a_\ell(k,\varepsilon)=\frac{1}{2\pi}\int_0^{2\pi}
d\theta\,\Delta\mathcal A(k,\theta,\varepsilon)e^{-i\ell\theta},
\quad P_\ell=|a_\ell|^2,
\label{eq:angular}
\end{equation}
where the integer \(\ell\) labels the angular harmonic. An \(h\)-wave response corresponds to \(|\ell|=5\), with ten alternating sign sectors around a momentum circle. As shown in Fig.~\ref{fig:spectrum}(c), $P_5$ dominates the displayed angular harmonics over the radial range with appreciable spin contrast, confirming the predominant $h$-wave character. Reversing the light helicity reverses the sign of the spin spectral contrast while leaving $P_\ell$ unchanged, demonstrating helicity-controlled reversal of the spin splitting without altering its $h$-wave symmetry (see SM Sec.~III~\cite{SuppMat} for details).

\Blue{\emph{Symmetry origin of light-induced $h$-wave magnetism.---}}
We first establish the undriven bulk spin degeneracy and then identify the symmetry constraints on the light-induced splitting. In quasicrystals, symmetry is defined through indistinguishability: symmetry-related configurations have identical bulk frequencies of finite local patterns, without requiring pointwise coincidence~\cite{Lifshitz1998, lifshitz1996symmetry, RevModPhys.63.699, RevModPhys.69.1181, Lifshitz2001QUASIPERIODICSS, Lifshitz2004SpinGroups,LIFSHITZ2005219}. For magnetic quasicrystals, these patterns include the magnetic labels and hopping amplitudes, rather than the tiling geometry alone. We denote independent spin and spatial operations by $[g_{\rm s}\Vert g_{\rm r}]$ and use $\overline{\mathcal A}_s$ for the thermodynamic bulk (or hull-averaged) spectral function. The resulting symmetry relations need not hold exactly for an arbitrary finite patch.

In the undriven system, the combined spin-flip--inversion operation $[C_{2\perp}^{\rm s}\Vert\mathcal P]$ relates indistinguishable magnetic configurations.
{Here $C_{2\perp}^{\rm s}$ denotes a $\pi$ spin rotation about an axis perpendicular to the N\'eel vector, while $\mathcal P$ is an affine hyperspace inversion, $\bm n\mapsto\bm e_\nu-\bm n$, which acts in physical space as $\bm r\mapsto\bm a_\nu-\bm r$, with $\bm a_\nu=S\bm e_\nu$. This operation exchanges the magnetic sublattices, while its perpendicular-space action preserves the bulk frequencies of magnetically labeled local patterns (see SM Sec.~IV~\cite{SuppMat}).}
The combined operation therefore relates opposite spins at opposite probe momenta. Moreover, each undriven spin block is real in the site basis, implying $\mathcal A_s(\kk,\varepsilon)=\mathcal A_s(-\kk,\varepsilon)$ through complex conjugation. Combining these relations gives
\begin{equation}
\overline{\mathcal A}_s(\kk,\varepsilon)
=\overline{\mathcal A}_{-s}(-\kk,\varepsilon)
=\overline{\mathcal A}_{-s}(\kk,\varepsilon).
\label{eq:undrivenbulk}
\end{equation}
Thus, the opposite-spin bulk spectra coincide at the same probe momentum, without requiring an exact time-spatial inversion symmetry or Kramers pairing of individual eigenstates in a finite Penrose patch.

Under CPL, the induced imaginary hopping removes the real-Hamiltonian constraint that enforces equal spectral weights at $\kk$ and $-\kk$ within each spin sector. Nevertheless, $[C_{2\perp}^{\rm s}\Vert\mathcal P]$ remains a bulk symmetry relation: $\mathcal P$ preserves the oriented angles and coherent path interference of the induced chiral hopping, while the accompanying spin rotation restores the staggered exchange field (see SM Sec.~IV~\cite{SuppMat}). Consequently,
\begin{align}
\overline{\mathcal A}_s(\kk,\varepsilon)
&=\overline{\mathcal A}_{-s}(-\kk,\varepsilon),\nonumber\\
\Delta_\lambda(-\kk)&=-\Delta_\lambda(\kk),
\label{eq:odd}
\end{align}
where $\Delta_\lambda(\kk)=E_{\lambda,\uparrow}(\kk)-E_{\lambda,\downarrow}(\kk)$ denotes the splitting between matched bulk spectral or quasi-band branches. Thus, the combined spin-flip--inversion symmetry enforces odd parity 
without requiring opposite-spin degeneracy at generic momentum.

The spin-preserving fivefold operation $[E_{\rm s}\Vert C_{5z}]$ further constrains the angular dependence:
\begin{equation}
\Delta_\lambda(k,\theta+2\pi/5)
=\Delta_\lambda(k,\theta).
\label{eq:selection}
\end{equation}
Expanding $\Delta_\lambda(k,\theta)$ in angular harmonics $e^{i\ell\theta}$, fivefold symmetry requires $\ell$ to be a multiple of five. Odd parity under $\theta\to\theta+\pi$ further restricts $\ell$ to odd integers, leaving $|\ell|=5,15,25,\ldots$. Thus, $p$- and $f$-wave components are forbidden, and $h$-wave is the lowest symmetry-allowed harmonic. For an analytic quasi-band branch near $\Gamma$, the lowest-order allowed contribution is
$\Delta_\lambda(k,\theta)
=k^5\left[c_\lambda\cos(5\theta)
+d_\lambda\sin(5\theta)\right]+O(k^7)$,
with coefficients determined by the microscopic hopping geometry.

Combining $[E_{\rm s}\Vert C_{5z}]$ with $[C_{2\perp}^{\rm s}\Vert\mathcal P]$ yields
\begin{equation}
\overline{\mathcal A}_s(R_{\pi/5}\kk,\varepsilon)
=\overline{\mathcal A}_{-s}(\kk,\varepsilon),
\label{eq:spinexchange}
\end{equation}
where $R_{\pi/5}$ rotates momentum by $\pi/5$. Accordingly, the spin-summed spectrum is invariant under a $\pi/5$ rotation, whereas the spin-difference spectrum changes sign, producing ten alternating spin sectors. This relation corresponds to the unitary spin-group operation $[C_{2\perp}^{\rm s}\Vert C_{10z}]$, rather than physical time-reversal symmetry~\cite{Luo2026HiddenZeeman}. Thus, optical driving permits spin splitting, while $[C_{2\perp}^{\rm s}\Vert\mathcal P]$ enforces odd parity and the fivefold rotational symmetry selects $h$-wave as the lowest allowed angular harmonic. The microscopic theory determines whether this splitting is nonzero, and its fifth-harmonic dominance is tested through Eq.~\eqref{eq:angular}.

\Blue{\emph{Long-wavelength quasi-band theory.---}}
The direct spectral calculation reveals the angular spin contrast without relying on an effective momentum-space model. To derive its microscopic origin and magnitude analytically, we construct a quasi-band theory by projecting the Hamiltonian onto states associated with groups of sites sharing specified local environments~\cite{wang2022effective}. In the cut-and-project construction, these groups correspond to subregions of the acceptance window~\cite{Koivusalo2021}. The effective hopping matrix elements are determined by overlaps of translated acceptance-window regions, which encode the frequencies of allowed hopping paths. For Floquet-induced two-step processes, these overlaps also enforce the existence of the intermediate vertex. Refining the window partition systematically incorporates more local environments. This construction provides an effective momentum-resolved description without assuming independent exact Bloch sectors, which are absent in a quasicrystal. The overlap weights and coherent spectral weights are derived in SM Sec.~V~\cite{SuppMat}.

For a group $W_b$ containing $N_b$ sites, we introduce the normalized state
\begin{equation}
|\kk,W_b,s\rangle=\frac{1}{\sqrt{N_b}}
\sum_{i\in W_b}e^{i\kk\cdot\rr_i}|i,s\rangle,
\label{eq:window}
\end{equation}
and construct the effective Hamiltonian from the matrix elements
$\langle\kk,W_a,s|H_{\rm F}|\kk,W_b,s\rangle$.
States associated with disjoint groups are orthogonal at fixed $\kk$, but generally not at different momenta. The resulting matrix elements retain coherent interference among allowed virtual-hopping paths.

At the coarsest level, retaining one state per magnetic sublattice yields
\begin{align}
\mathcal H^{(0)}(\kk)&=N\sigma_z\tau_z-B_5(\kk)\tau_x
-\eta D_5(\kk)\tau_z,
\label{eq:Hk}
\end{align}
where $\bm\tau$ are Pauli matrices in the $A/B$ sublattice space and $\bm a_j=a(\cos 2\pi j/5,\sin 2\pi j/5)$ denote the five edge directions. The even function $B_5={4t_1J_0(\alpha_0)}
\sum_{j=0}^4\cos(\kk\cdot\bm a_j)/5$ describes real inter-sublattice hybridization, whereas the chiral intra-sublattice hopping generates opposite diagonal shifts on the two sublattices, characterized by the odd function
\begin{align}
D_5(\kk)&=\frac{\xi}{5}
\sum_{\delta=2\pi/5,\,4\pi/5}
\Lambda(\delta)\sum_{j=0}^4
\sin(L_\delta\kk\cdot\bm u_j).
\label{eq:D5}
\end{align}
Here the geometric factor $\xi=5\sqrt5-11$ arises from the local-environment-weighted path counts. The two values of $\delta$ correspond to the surviving virtual-hopping geometries, with endpoint separations $L_\delta=2a\sin(\delta/2)$ and displacement directions $\bm u_j=(\cos[(4j+1)\pi/10],\sin[(4j+1)\pi/10])$.
Thus, $D_5$ follows directly from the Floquet-induced virtual hopping in Eq.~\eqref{eq:H2}, rather than being introduced phenomenologically.

Diagonalizing the effective Hamiltonian gives
\begin{equation}
E_{\lambda s}(\kk)=
\lambda\sqrt{B_5^2(\kk)+[sN-\eta D_5(\kk)]^2},
\label{eq:bands}
\end{equation}
where $\lambda=\pm1$ labels the upper and lower quasi-bands for each spin sector. The cross term $-2s\eta ND_5$ reveals the cooperative origin of the splitting: both staggered exchange and chiral hopping are required. For $|D_5|\ll E_0=\sqrt{N^2+B_5^2}$, the spin splitting becomes $\Delta_\lambda=-2\lambda\eta ND_5/E_0+O(D_5^3/E_0^2)$.
Reversing the light helicity exchanges the opposite-spin quasi-band energies while leaving their spin average unchanged.

In the long-wavelength limit $ka\ll1$, the fivefold sum in Eq.~\eqref{eq:D5} cancels the linear and cubic terms, leaving a leading fifth-order contribution (see SM Sec.~VI~\cite{SuppMat}). The resulting spin splitting is
\begin{equation}
\Delta_\lambda(\kk)=
-\frac{2\lambda\eta N g_5}
{\sqrt{N^2+16t_1^2J_0^2(\alpha_0)}}
k^5\sin(5\theta)+O(k^7),
\label{eq:h}
\end{equation}
where
$g_5=(5\sqrt5-11)\sum_{\delta=2\pi/5,\,4\pi/5} \Lambda(\delta)L_\delta^5/1920$.
The $\sin(5\theta)$ dependence fixes the nodal orientation through the virtual-hopping geometry and divides the momentum plane into ten alternating spin sectors. Equation~\eqref{eq:h} thus provides a microscopic derivation of the symmetry-allowed $h$-wave splitting, accounting for the angular pattern in Fig.~\ref{fig:spectrum}(b) and its reversal under opposite light helicities.

\begin{figure}[t]
\centering
\includegraphics[width=1\columnwidth]{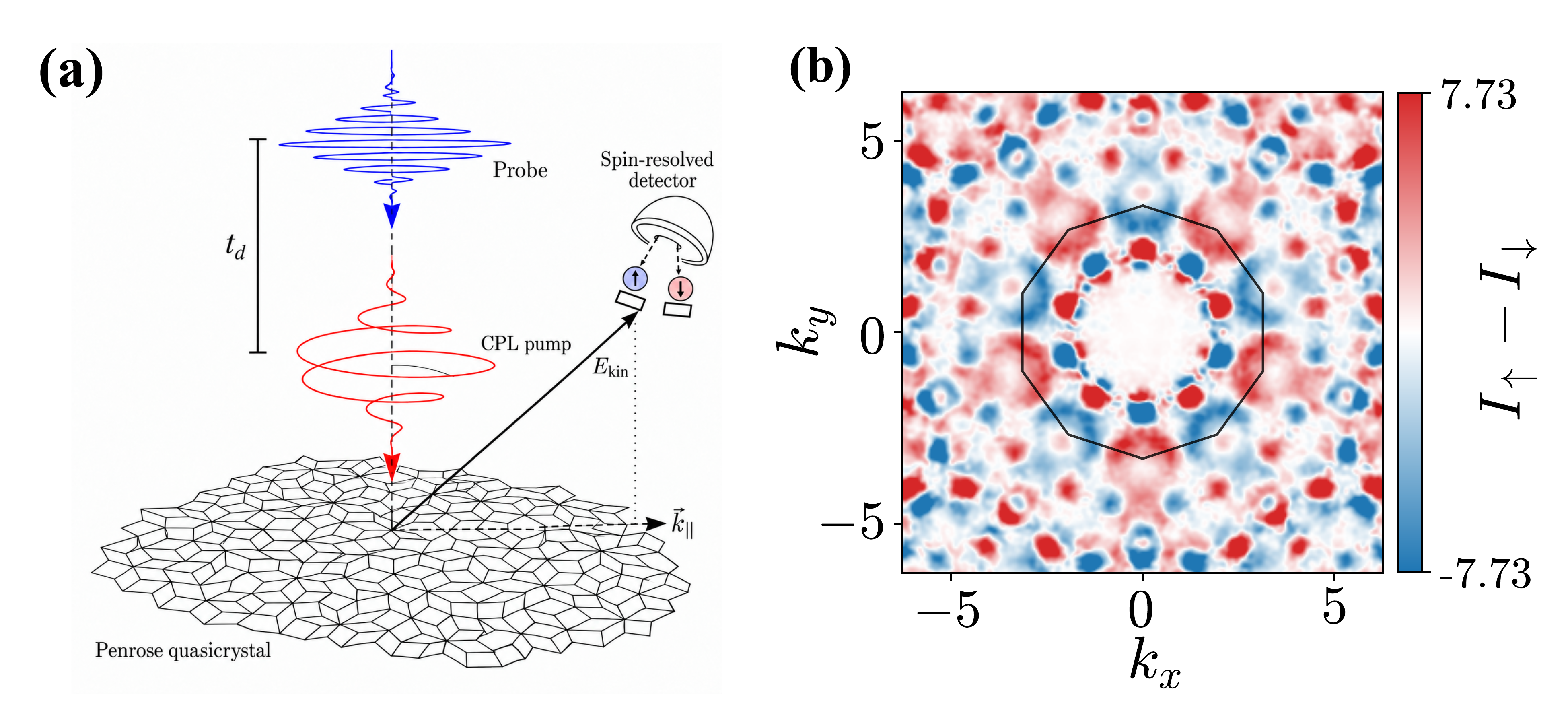}
\caption{\label{fig:optical}
Spin-resolved pump--probe photoemission.
(a) Schematic of the detection protocol: a circularly polarized pump drives the sample, followed by a delayed probe that resolves the spin and in-plane momentum of photoemitted electrons.
(b) Calculated spin contrast $I_\uparrow-I_\downarrow$ at $\varepsilon=-1.20\,{\rm eV}$ and delay $t_d=0$.
Parameters are $t_1=N=1\,{\rm eV}$, $\alpha_0=1.5$, $\omega=20\,{\rm rad/fs}$, $T_{\rm pump}=150\,{\rm fs}$, $T_{\rm probe}=10\,{\rm fs}$, $T=0$, and $\mu=0$.
}
\end{figure}

\Blue{\emph{Spin-resolved trARPES and optical control.---}}
To probe the light-induced $h$-wave spin splitting, we consider spin-resolved trARPES in a pump--probe setup [Fig.~\ref{fig:optical}(a)]~\cite{PhysRevLett.102.136401, Sentef2015, PhysRevB.94.155304, PhysRevResearch.4.013057, RevModPhys.96.015003}. To account for the finite pulse duration and probe resolution, we employ a circularly polarized pump
$\bA_{\rm pump}(t)=A_0e^{-(t/T_{\rm pump})^2}(\cos\omega t,\eta\sin\omega t)$
and a Gaussian probe envelope
$s_{\rm pr}(t,t_d)=e^{-[(t-t_d)/T_{\rm probe}]^2}$,
where $T_{\rm pump}$ and $T_{\rm probe}$ are the $1/e$ half-widths of the respective amplitude envelopes, and $t_d$ is the probe delay relative to the pump maximum. Neglecting momentum-dependent photoemission matrix elements and final-state effects, the spin-resolved photocurrent is given by~\cite{PhysRevLett.102.136401,Sentef2015}
\begin{multline}
I_s(\kk,\varepsilon,t_d)\propto\operatorname{Im}
\int dt_1dt_2\,s_{\rm pr}(t_1,t_d)s_{\rm pr}(t_2,t_d)\\
{}\times e^{i\varepsilon(t_1-t_2)/\hbar}
G^<_{ss}(\kk;t_1,t_2),
\label{eq:arpes}
\end{multline}
where $G^<_{ss}(\kk;t_1,t_2)
=i\langle c^\dagger_{\kk s}(t_2)c_{\kk s}(t_1)\rangle$
is the lesser Green function describing occupied electronic correlations. Here $c_{\kk s}$ is the normalized plane-wave transform of the site operators, as defined in Eq.~\eqref{eq:spectral}. The probe envelope selects a time window centered at $t_d$, while the Fourier phase resolves the photoemission energy $\varepsilon$.

For independent electrons initially described by $H_{\rm init}$, the real-space lesser Green function evolves as
\begin{equation}
G^<(t_1,t_2)=iU(t_1,t_0)f(H_{\rm init})U^\dagger(t_2,t_0),
\label{eq:propagatedlesser}
\end{equation}
where $U(t,t_0)$ is the single-particle evolution operator under the pump and $f(H_{\rm init})$ is the initial Fermi--Dirac occupation matrix, with chemical potential $\mu$ and temperature $T$. Projecting this correlator onto plane-wave states yields $G^<_{ss}(\kk;t_1,t_2)$ in Eq.~\eqref{eq:arpes}. Details of the derivation and numerical implementation are provided in SM Sec.~VII~\cite{SuppMat}.

The calculated spin-difference photocurrent $I_\uparrow-I_\downarrow$ [Fig.~\ref{fig:optical}(b)] exhibits ten alternating angular sectors, superimposed on diffraction-related intensity modulations. Reversing the CPL helicity interchanges the opposite-spin photocurrents, providing an additional means of optical control over the spectroscopic spin contrast.
Its sign indicates the dominant spin contribution to the occupied photoemission weight at a given energy and probe delay. The resulting pattern provides a spin-resolved spectroscopic signature of the predicted $h$-wave electronic structure. Unlike the Floquet spectral function, however, the photocurrent also depends on nonequilibrium occupations, pulse envelopes, and probe resolution. These effects, together with the relevant optical scales and experimental limitations, are discussed in SM Sec.~VIII~\cite{SuppMat}.

\Blue{\emph{Discussion and conclusions.---}}
We have demonstrated that CPL can induce odd-parity $h$-wave spin splitting in an antiferromagnetic Penrose quasicrystal without SOC. The central conceptual advance lies in the complementary roles of optical driving and noncrystallographic rotational symmetry: the former generates the orbital hopping responsible for lifting spin degeneracy, while the latter, together with spin-spatial symmetry, selects the unconventional $h$-wave angular structure. This mechanism extends symmetry-guided design of momentum-dependent spin splitting beyond periodic crystals, where crystallographic restrictions exclude fivefold rotational symmetry.

Our predictions concern the driven electronic structure on a fixed antiferromagnetic background, rather than the formation of a new magnetic order. Their experimental realization will depend on material-specific optical coupling, nonequilibrium occupations, heating, and relaxation, while self-consistent magnetic dynamics remain an important direction for future study. More broadly, the interplay between optical driving and noncrystallographic symmetry offers a route to engineering unconventional spin-split electronic structures in quasiperiodic and other aperiodic magnetic systems.

\begin{acknowledgments}
This work is supported by the National Natural Science Foundation of China (Grants No. 12474056 and No. 12674195), the National Key R\&D Program of China (Grant No. 2021YFA1401600), and the 2022 basic discipline top-notch students training program 2.0 research project (Grant No. 20222005). The work was carried out at the National Supercomputer Center in Tianjin, and the calculations were performed on Tianhe new generation supercomputer. The high-performance computing platform of Peking University supported the computational resources.
\end{acknowledgments}


%

\appendix

\section{Bipartite Penrose tiling: cut-and-project construction}
\label{sec:bipartite}

\subsection{Projection from five dimensions}

We construct the Penrose tiling from the integer lattice
\(\bm n=(n_1,\ldots,n_5)^T\in\mathbb Z^5\), {with standard hypercubic basis \(\bm e_\nu\), \(\nu=1,\ldots,5\)}.
The five-dimensional space is decomposed into a two-dimensional physical
space \(V_\parallel\) and a three-dimensional perpendicular space
\(V_\perp\).  A convenient choice of projection matrices is
\begin{align}
{S}=\frac{1}{2}
 \begin{pmatrix}
 2&-\sigma&-\tau&-\tau&-\sigma\\
 0&\beta\tau&\beta&-\beta&-\beta\tau
 \end{pmatrix},                                      \label{Sproj}\\
{S_\perp}=\frac{1}{2}
 \begin{pmatrix}
 2&-\tau&-\sigma&-\sigma&-\tau\\
 0&\beta&-\beta\tau&\beta\tau&-\beta\\
 \sqrt2&\sqrt2&\sqrt2&\sqrt2&\sqrt2
 \end{pmatrix},                                      \label{Sperp}
\end{align}
where
\(\tau=(1+\sqrt5)/2\), \(\sigma=(1-\sqrt5)/2=-1/\tau\),
and \(\beta=\sqrt{3-\tau}\).
Each hyperlattice site has physical and perpendicular coordinates
\[
 {\rr(\bm n)}={S}\bm n,\qquad
 {\rr_\perp(\bm n)}={S_\perp}\bm n.
\]
{
The displayed coordinates use the edge length as the unit, \(a=1\); physical lengths are restored by multiplying its entries by \(a\).
The columns \(S\bm e_\nu\), \(\nu=1,\ldots,5\), give the same five edge directions as the main-text star \(\bm a_j\), with \(j=\nu-1\). Noting that the geometric constants \(\sigma,\tau\) in these matrices are scalars, distinct from the spin and sublattice Pauli matrices \(\bm\sigma,\bm\tau\).}

A site is retained when its perpendicular coordinate lies inside the
acceptance window \(W_\perp\).  The window is the projection of the
five-dimensional unit hypercube, {up to the fixed internal-space
shift that specifies the layer origin below}, and is a three-dimensional polytope with
decagonal cross sections.  One representation of its vertices in
cylindrical coordinates consists of
\begin{align*}
 &(0,0,3/\sqrt2),\quad
 (1,\theta_1,2/\sqrt2),\quad
 (\tau,\theta_2,1/\sqrt2),\\
 &(\tau,\theta_1,0),\quad
 (1,\theta_2,-1/\sqrt2),\quad
 (0,0,-2/\sqrt2),
\end{align*}
with
\[
 \theta_1\in\{\pm\pi/5,\pm3\pi/5,\pi\},\qquad
 \theta_2\in\{0,\pm2\pi/5,\pm4\pi/5\}.
\]
The finite samples used in numerical calculations are obtained by
restricting \(\bm n\) to a sufficiently large hypercubic box, applying
the window test, and then selecting a nearly circular patch in physical
space.  Boundary sites can either be retained with open boundaries or
discarded from local observables using an interior mask.

\subsection{Magnetic sublattices and inversion}

The parent hypercubic lattice is bipartite.  We assign
\[
 \chi(\bm n)=(-1)^{n_1+n_2+n_3+n_4+n_5},
\]
and define sublattice \(A\) by \(\chi=+1\) and \(B\) by \(\chi=-1\).
A step along any hypercubic basis vector changes \(\chi\), hence every
projected nearest-neighbor bond connects opposite sublattices.  The
staggered exchange term is therefore
\[
 {H_N=N\sum_i\chi_i c_i^\dagger\sigma_z c_i.}=N\sum_i c_i^\dagger\sigma_z\tau_z c_i
\]
Here \(c_i\) is a two-component spinor. The sublattice Pauli matrix \(\tau_z\) is introduced only after projection onto the explicit \(A/B\) basis, with \(\tau_z=+1\) on \(A\) and \(-1\) on \(B\).
Pure parent-lattice inversion \(\bm n\to-\bm n\) preserves \(\chi\) and does not exchange \(A\) and \(B\). The sublattice-exchanging operation denoted by \(\mathcal P\) is instead realized by the affine map \(\bm n\mapsto\bm e_\nu-\bm n\), where \(\bm e_\nu\) is a parent-lattice basis vector. It acts in physical space as inversion about \({S}\bm e_\nu/2\) and maps the acceptance window to a common transverse internal-space translate. Section~\ref{sec:symmetry} proves that this preserves the bulk frequencies of magnetically labeled hopping patterns for the regular window specified here. It does not require an arbitrary finite patch to coincide with its image.

\subsection{Local environments}

\begin{figure}[t]
 \centering
 \includegraphics[width=0.95\linewidth]{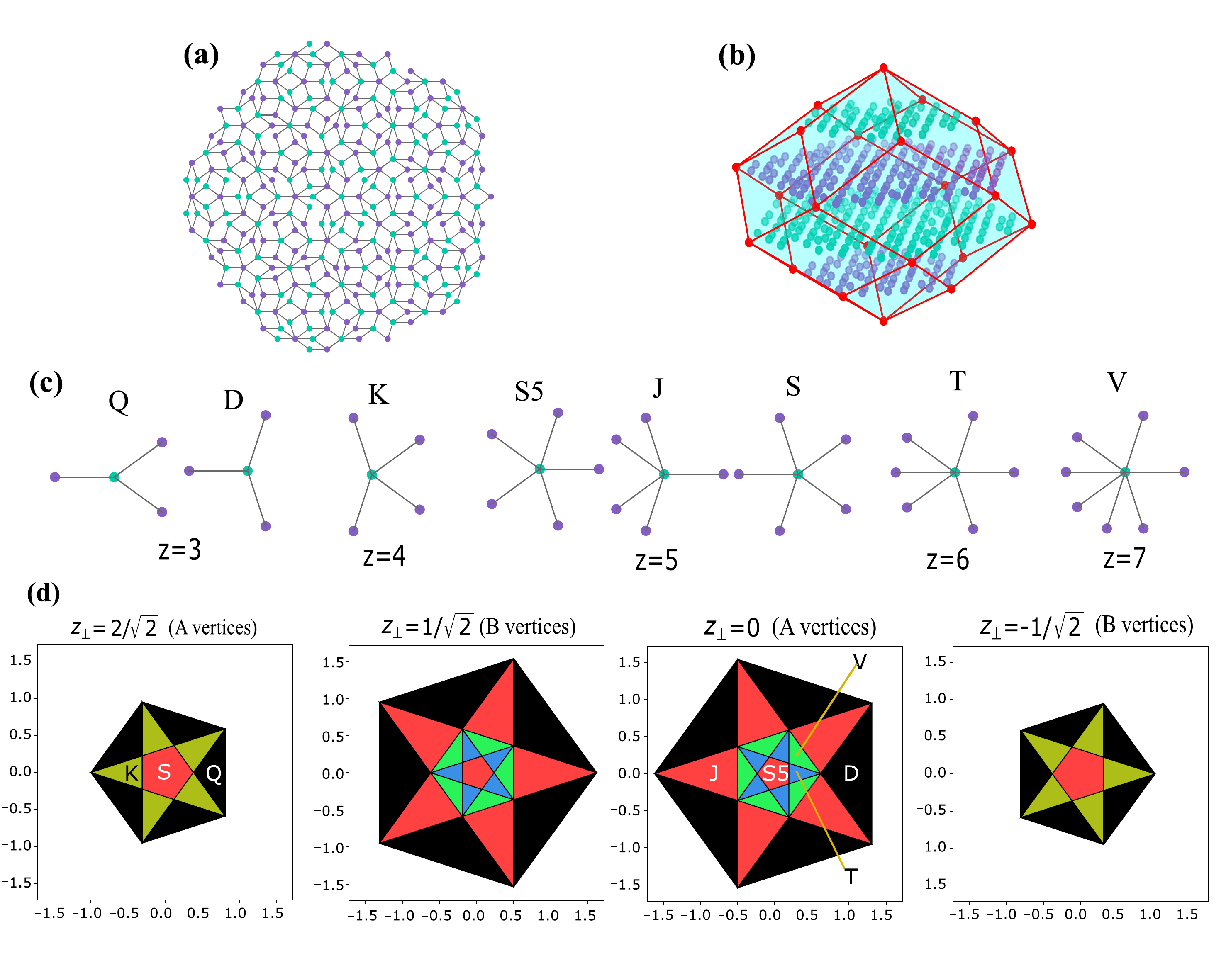}
 \caption{\label{fig:S1}Cut-and-project construction and local
 classification of the antiferromagnetic Penrose tiling.
 The illustrative patch has 301 vertices. The physical and perpendicular
 projections and the eight local environments illustrate the construction.}
\end{figure}

The first-neighbor coordination number ranges from \(z=3\) to \(z=7\).
Accounting also for bond orientation yields eight vertex classes,
traditionally denoted \(D,Q,K,S,S5,J,T,V\) \cite{DEBRUIJN198153, PhysRevB.34.6924, PhysRevB.40.11185}.  Each class corresponds to
a subregion of \(W_\perp\), and its thermodynamic frequency is the
induced measure of that subregion divided by the total occupied-window
measure. For the four-layer Penrose construction, this is a sum of
polygonal areas over the occupied layers, not an unrestricted integral
through the three-dimensional convex hull.
The weights used in the two-step enumeration are
\begin{align}
\rho_Q&=-\frac{\sigma}{\tau^3},&
\rho_D&=-\frac{\sigma}{\tau},&
\rho_K&=\frac{\sigma^2}{\tau^3},\nonumber\\
\rho_S&=\frac{\sigma^2}{\sqrt5\,\tau^3},&
\rho_{S5}&=\frac{\sigma^4}{\sqrt5\,\tau^3},&
\rho_J&=\frac{\sigma^2}{\tau},\nonumber\\
\rho_T&=\frac{\sigma^4}{\tau^3},&
\rho_V&=-\frac{\sigma^3}{\tau^3}.            \label{vertexweights}
\end{align}
The construction and perpendicular-space subwindows are shown in Fig.~\ref{fig:S1}. All rotated versions of a local environment belong to the same class.

\section{High-frequency Floquet expansion}
\label{sec:floquet}

\subsection{Fourier components of the Peierls hopping}

For a directed nearest-neighbor bond \(j\to i\), let \(\rr_{ij}=a(\cos\theta_{ij},\sin\theta_{ij})\). Here \(\theta_{ij}\) is measured counterclockwise from the positive \(x\) axis to \(\rr_i-\rr_j\). The orbital coupling to a spatially uniform light field is included by the Peierls phase \(\exp[i(e/\hbar)\bm A(t)\cdot\rr_{ij}]\), with \(e>0\). For CPL, \(\bA(t)=A_0(\cos\omega t,\eta\sin\omega t)\), with light helicity \(\eta=\pm1\) for the fixed propagation direction used in the main text. Then
\begin{align}
 \frac{e}{\hbar}\bA(t)\cdot\rr_{ij}
 &=\alpha_0[
 \cos\omega t\cos\theta_{ij}
 +\eta\sin\omega t\sin\theta_{ij}]\nonumber\\
 &=\alpha_0\cos(\omega t-\eta\theta_{ij}),
\end{align}
where \(\alpha_0=eA_0a/\hbar\).  The Jacobi--Anger identity gives
\[
 e^{i\alpha_0\cos(\omega t-\eta\theta)}
 =\sum_{n=-\infty}^{\infty}
 i^nJ_n(\alpha_0)e^{-in\omega t}e^{in\eta\theta}.
\]
Here \(J_n\) is the Bessel function of the first kind of integer order
\(n\), including the zeroth-order function \(J_0\).
With \(H(t)=\sum_nH_ne^{-in\omega t}\), the Fourier components are
\[
H_n=\frac{1}{T_{\rm L}}\int_0^{T_{\rm L}}dt\,H(t)e^{in\omega t},
\]
where \(T_{\rm L}=2\pi/\omega\). For \(n\ne0\), only hopping contributes,
and therefore
\begin{equation}
 H_n=-t_1i^nJ_n(\alpha_0)
 \sum_{\langle ij\rangle}
 e^{in\eta\theta_{ij}}c_i^\dagger c_j .
 \label{Hn}
\end{equation}
The time average \(H_0\) renormalizes the real hopping by
\(J_0(\alpha_0)\) and leaves the staggered exchange unchanged.
The sum contains both bond orientations, so \(H_{-n}=H_n^\dagger\).
Using the opposite Fourier convention exchanges the order of the
commutator in the effective Hamiltonian. Keeping these conventions
together is essential for a consistent helicity sign.

\subsection{Commutator and induced same-sublattice hopping}

To first order in \(1/\omega\),
\[
 H_{\rm F}=H_0+\sum_{n>0}\frac{[H_{-n},H_n]}{n\hbar\omega}+O(\omega^{-2}).
\]
This is the van Vleck Hamiltonian \cite{Eckardt2015}. To the retained order the evolution
has the form \(U(t,t_0)=e^{-iK(t)}
e^{-iH_{\rm F}(t-t_0)/\hbar}e^{iK(t_0)}\), with a periodic kick
\(K(t)\). A stroboscopic Hamiltonian at a specified time origin is
unitarily related to this operator; the kick matters for spectral weights.
Terms in the commutator sharing an intermediate site generate
\(i\to j\to k\).  Combining a path with its Hermitian conjugate yields
\begin{align}
 H^{i-j-k}_{\rm F}&=i\eta\Lambda(\delta)
 (c_i^\dagger c_k-c_k^\dagger c_i),\label{twoStep}\\
 \Lambda(\delta)&=\frac{2t_1^2}{\hbar\omega}
 \sum_{n>0}\frac{J_n^2(\alpha_0)}{n}\sin(n\delta).
 \label{LambdaSM}
\end{align}
Equation~\eqref{twoStep} is the effective hopping between endpoints
\(i\) and \(k\) mediated by the intermediate site \(j\). Its angle
\(\delta=\theta_{kj}-\theta_{ij}\) is the oriented difference between
the two bonds pointing outward from \(j\). Each nearest-neighbor hop
changes magnetic sublattice, so the two endpoints belong to the same
sublattice. Equation~\eqref{twoStep} has the
following properties:
\begin{enumerate}
 \item it is imaginary and antisymmetric under \(i\leftrightarrow k\);
 \item it changes sign when the helicity \(\eta\) is reversed;
 \item it vanishes for a collinear two-step path;
 \item it is odd under reversal of the oriented turning angle;
 \item its leading weak-field scale is
 \(t_1^2\alpha_0^2/(\hbar\omega)\).
\end{enumerate}

The same basic cooperation between a spin-independent orbital term and staggered exchange appears in crystalline Floquet antiferromagnets
\cite{Wangrui2026_PRL,Ruanjiawei2026_PRL}. Lin and Vila formulate a related mechanism in terms of opposite sublattice currents: the orbital sector produces a momentum-dependent sublattice imbalance that exchange converts into spin splitting \cite{LinVila2026Sublattice}. Our calculation fixes the exchange field and derives the imaginary hopping from the external drive. A complex hopping coefficient is not itself an expectation value
of a bond-current operator; neither spontaneous loop-current order nor a steady current distribution is calculated here. The specifically
quasicrystalline input is the coherent sum of geometrically allowed paths.

If \(i\) and \(k\) are opposite vertices of a Penrose rhombus, there are two intermediate vertices \(j_1,j_2\).  The two paths enclose opposite
oriented areas and have opposite turning angles.  Their amplitudes in Eq.~\eqref{twoStep} cancel.  Removing these diagonal pairs and
enumerating the remaining paths around the eight vertex classes gives 0, 0, 2, 5, 5, 5, 8, and 12 directed processes around \(D,Q,K,S,S5,J,T,V\), respectively\cite{PhysRevB.34.6924, PhysRevB.40.11185}.

\begin{figure}[t]
 \centering
 \includegraphics[width=0.95\linewidth]{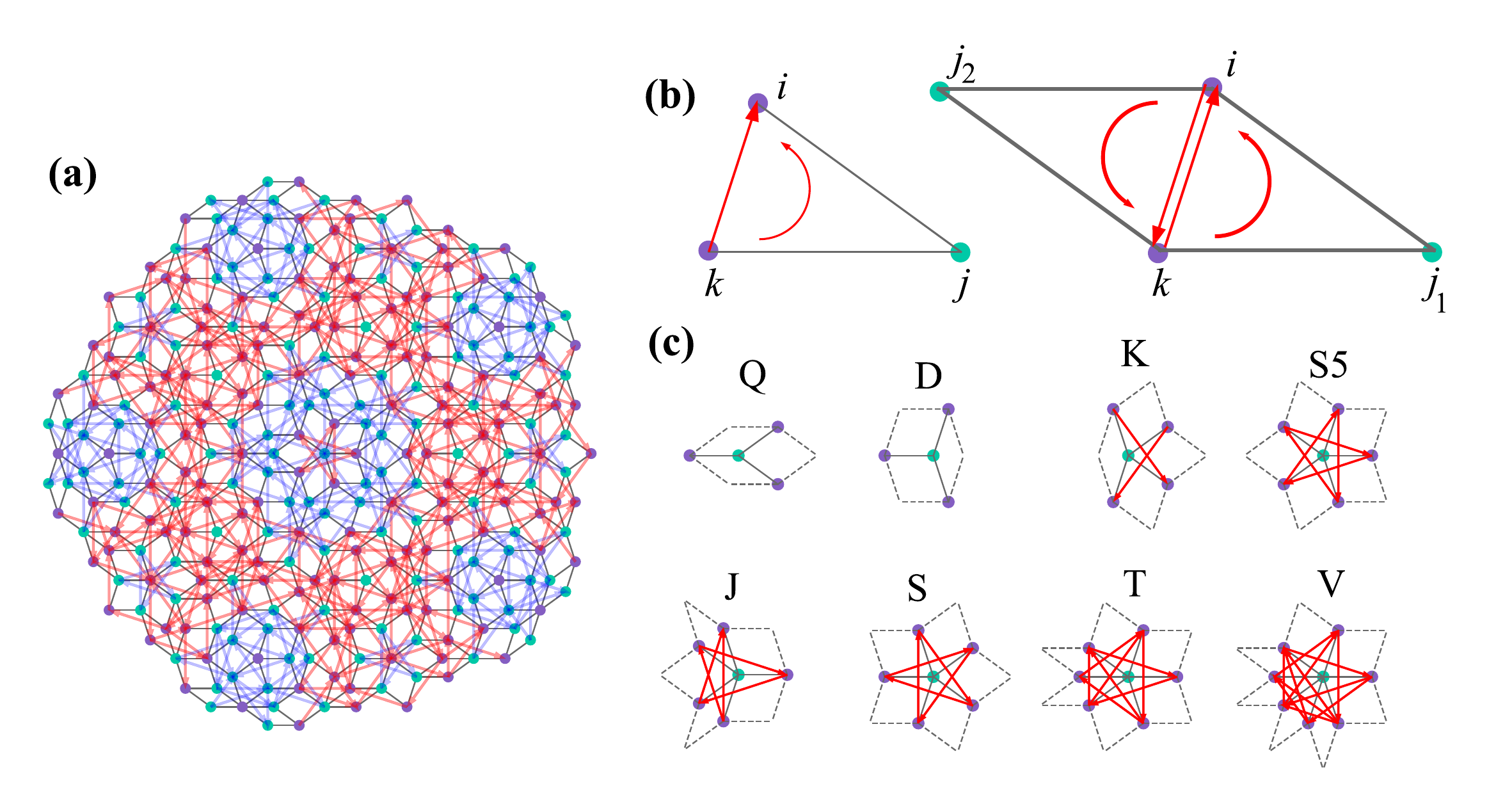}
 \caption{\label{fig:S2}Enumeration of Floquet-induced hopping.
 (a) Same-sublattice hopping in the illustrative 301-site patch.
 (b) Orientation convention and rhombus cancellation.
 (c) Remaining paths around each vertex class, with \(B\)--\(B\)
 hopping shown as an example.}
\end{figure}

\section{Finite-tiling spectra and numerical validation}
\label{sec:numericsSM}

We construct the finite real-space effective Hamiltonian before applying the local-environment projection. Plane-wave overlaps resolve its spectral weight at continuous probe momentum, while a circular angular decomposition characterizes the spin contrast.
For Fig~2 of the main text, we use parameters: \(N_{\rm site}=1387\), \(t_1=N=1\,{\rm eV}\), \(\omega=20\,{\rm rad/fs}\), \(\alpha_0=1.5\), and \(\Gamma=0.01\,{\rm eV}\). The fixed-energy map and its angular analysis are evaluated at \(\varepsilon=-1.20\,{\rm eV}\). The Hamiltonian assembled below retains terms through order \(1/\omega\).

\subsection{Effective Hamiltonian and path bookkeeping}
For each spin block, assemble
\begin{equation}
{H_{{\rm F},s}}
=sN\sum_i\chi_i|i\rangle\langle i|
-t_1J_0(\alpha_0)\sum_{\langle ij\rangle}|i\rangle\langle j|
+\sum_j\sum_{\{i,k\}\subset{\cal N}(j)}
i\eta\Lambda(\theta_{kj}-\theta_{ij})
(|i\rangle\langle k|-|k\rangle\langle i|).
\label{realHF}
\end{equation}
Here \(H_{{\rm F},s}\) is the spin-\(s\) block of the real-space effective Hamiltonian \(H_{\rm F}\) in the main text. 
\({\cal N}(j)\) is the nearest-neighbor list of \(j\).
The edge sum contains both orientations, whereas the last sum contains
each unordered pair of distinct neighbors only once. Choose either
endpoint ordering for that pair and use it consistently in both the
angle and the operator. Exchanging \(i,k\) reverses both signs and leaves
the contribution unchanged. This convention avoids an extra factor
of two. Contributions from different intermediate sites are accumulated
on the same endpoint matrix element; rhombus cancellations then occur
algebraically, not by removing arbitrary endpoint pairs.

The Bessel sum can be truncated at \(n_{\max}\). Its accuracy is characterized by the change in matrix elements and spectra upon increasing \(n_{\max}\). Direct path assembly and the Fourier-commutator construction give equivalent matrices when the same bond orientations and cutoffs are used. Hermiticity is measured by \(\|H-H^\dagger\|_{\rm F}/\|H\|_{\rm F}\), with \(\|\cdot\|_{\rm F}\) the Frobenius norm. Retaining only \(n=1\) gives the leading weak-drive approximation. For the amplitude of \(\alpha_0=1.5\) in the main text, this truncation is set as $n_{\max}=8$.

\subsection{Plane-wave projection, normalization, and spectral cuts}
With real radial weights \(w_i\), define a normalized probe state
\begin{equation}
|\kk,s;w\rangle=\frac{1}{\sqrt{\sum_iw_i^2}}
 \sum_iw_i e^{i\kk\cdot\rr_i}|i,s\rangle .
\label{probeSM}
\end{equation}
The main text uses \(w_i=1\). A smooth mask can suppress the outer
boundary. The mask affects boundary sensitivity and angular resolution; a symmetry-breaking mask can produce forbidden angular components.

The spectral function is
\begin{align}
\mathcal A_s(\kk,\varepsilon)
&=-\frac{1}{\pi}\operatorname{Im}
\langle\kk,s;w|(\varepsilon+i\Gamma-{H_{{\rm F},s}})^{-1}
|\kk,s;w\rangle \label{resolvent}\\
&=\sum_m|\langle\kk,s;w|m,s\rangle|^2
 \frac{\Gamma/\pi}{(\varepsilon-E_{ms})^2+\Gamma^2}.
\label{spectralsumSM}
\end{align}
For a complete eigenbasis, \(\mathcal A_s\geq0\) and
\(\int_{-\infty}^{\infty}\mathcal A_s\,d\varepsilon=1\) at each \(\kk\).
An integral on a finite numerical energy grid misses the Lorentzian
tails; its truncation error must not be mistaken for loss of state
normalization. The real-space eigenvalues themselves have no momentum
label. Dispersive peaks arise from their momentum-dependent overlaps.

Using the same patch, mask, and broadening for both spins isolates the spin difference \(\Delta\mathcal A=\mathcal A_\uparrow-\mathcal A_\downarrow\). The spin sum identifies regions of appreciable spectral intensity. A large normalized polarization where this sum vanishes has little spectroscopic weight. The undriven spectrum provides a baseline for finite-boundary effects.

\subsection{Complex angular harmonics and fifth-component weight}
For a circle of fixed radius \(k\) and fixed energy \(\varepsilon\),
sample \(\theta_j=2\pi j/N_\theta\), \(j=0,\ldots,N_\theta-1\), and use
\begin{align}
a_\ell(k,\varepsilon)&=\frac{1}{N_\theta}
 \sum_j\Delta\mathcal A(k,\theta_j,\varepsilon)e^{-i\ell\theta_j},
\label{complexangularSM}\\
P_\ell(k,\varepsilon)&=|a_\ell(k,\varepsilon)|^2 .
\end{align}
This is the discrete form of the main-text definition. For real
\(\Delta\mathcal A\), \(a_{-\ell}=a_\ell^*\). If real coefficients are desired,
define \(c_\ell=2\operatorname{Re}a_\ell\) and
\(d_\ell=-2\operatorname{Im}a_\ell\) for \(\ell>0\), so that
\(\Delta\mathcal A=a_0+\sum_{\ell>0}(c_\ell\cos\ell\theta+
d_\ell\sin\ell\theta)\).
The amplitude \(W_\ell=\sqrt{c_\ell^2+d_\ell^2}=2|a_\ell|\)
therefore obeys \(P_\ell=W_\ell^2/4\).
This fixes the factor-of-two convention that differs from some
real sine/cosine decompositions. For a pure \(F\sin5\theta\),
\(a_5=-iF/2\) and \(P_5=F^2/4\).

Angular sampling with \(N_\theta>2\ell_{\max}\) resolves harmonics through \(\ell_{\max}\); refinement controls aliasing from higher components. A normalized fifth-harmonic fraction is
\begin{equation}
f_5(k,\varepsilon)=
\frac{P_5}{\sum_{\ell=1}^{\ell_{\max}}P_\ell},
\label{fractionSM}
\end{equation}
defined when the denominator is nonzero. The absolute weight \(P_5\) and the retained cutoff distinguish a strong signal from a large fraction of a weak signal. Even and symmetry-forbidden components diagnose boundary or discretization effects. Figure~2(c) in the main text compares the displayed angular components; the bulk symmetry also allows \(\ell=15,25,\ldots\), whose weights are not fixed by symmetry. 

The complex coefficient $a_\ell$, unlike \(P_\ell\), retains the sign and angular
phase of the spin contrast. To state the helicity relation explicitly,
write \(H_{{\rm F},s}^{(\eta)}\) and
\(\mathcal A_s^{(\eta)}\) for the effective Hamiltonian and spectrum
at helicity \(\eta\). Complex conjugation in the site basis gives
\begin{equation}
K H_{{\rm F},s}^{(\eta)} K^{-1}=H_{{\rm F},s}^{(-\eta)},
\qquad
\mathcal A_s^{(-\eta)}(\kk,\varepsilon)
=\mathcal A_s^{(\eta)}(-\kk,\varepsilon).
\label{helicityConjugationSM}
\end{equation}
The latter identity holds for a finite patch with the same real mask
and broadening. Combining it with the bulk combined spin-flip--inversion
relation in Sec.~\ref{sec:symmetry} gives
\begin{equation}
\overline{\mathcal A}_s^{(-\eta)}(\kk,\varepsilon)
=\overline{\mathcal A}_{-s}^{(\eta)}(\kk,\varepsilon).
\label{bulkHelicitySM}
\end{equation}
Thus the bulk spin contrast and its complex angular coefficients
reverse sign, while their squared weights are unchanged:
\(a_\ell^{(-\eta)}=-a_\ell^{(\eta)}\) and
\(P_\ell^{(-\eta)}=P_\ell^{(\eta)}\).
These statements concern the Floquet spectrum; finite boundaries can modify the bulk relations. 

\subsection{Extracting an energy splitting and testing the radial law}
A fixed-energy \(\Delta\mathcal A\) is not an energy difference. For example, for two identical line shapes displaced by
\(\Delta/2\) in opposite directions, {\(\Delta\mathcal A\simeq-\Delta\,\partial_\varepsilon\mathcal A_{\rm ref}\), where \(\mathcal A_{\rm ref}\) is the common unsplit line shape}, to first order. Changes of spectral weight add further terms. Thus a spectral harmonic can change sign with energy even when the underlying splitting does not.

An energy splitting is extracted by tracking matched, resolvable opposite-spin peaks continuously in \(k\) and \(\theta\). Denoting the difference by \(\Delta_\lambda^{\rm peak}\), its fifth harmonic is
\begin{equation}
{a_{\lambda,5}^{\rm peak}}(k)=\frac{1}{2\pi}
\int_0^{2\pi}d\theta\,
\Delta_\lambda^{\rm peak}(k,\theta)e^{-5i\theta}.
\label{peakangularSM}
\end{equation}
In the coordinate convention of Sec.~\ref{sec:quasi-band}, the analytic prediction is \(-iK_\lambda k^5/2\), where
\(K_\lambda=-2\lambda\eta Ng_5/\sqrt{N^2+16t_1^2J_0^2(\alpha_0)}\).
The phase distinguishes orientation and sign, while the modulus measures the splitting amplitude. A leading radial correction has the form \(Ck^5(1+bk^2)\). Overlapping or unresolved peaks do not define an unambiguous branch splitting; {this energy-based diagnostic is separate from the fixed-energy intensity weights shown in main-text Fig.~2(c), and those weights alone do not establish a numerical radial \(k^5\) law.}

\subsection{Controls and finite-size convergence}
The microscopic mechanism has three null limits: no illumination, no staggered exchange, and omission of the chiral commutator. In each, the bulk spin splitting vanishes. Reversing circular helicity changes the sign of the chiral hopping. For a monochromatic linearly polarized drive, the leading chiral commutator vanishes because there is no oriented optical area. 

This control should not be generalized to arbitrary combinations of linearly polarized fields. Yu considers orthogonal components at frequencies \(\omega\) and \(2\omega\), with their relative phase chosen to generate a distinct orbital symmetry breaking \cite{Yu2026MixedParity}. For the time-even waveform used there, \(H_n=H_{-n}\), so the order-\(1/\omega\) commutators vanish; an odd-parity contribution nevertheless arises from the cycle-averaged Hamiltonian in an initially even-parity altermagnet. The resulting splitting has mixed parity. This differs both in starting state and in Floquet order from our CPL-induced splitting of a spin-degenerate antiferromagnet. 

Finite-sample departures from the bulk symmetry relations can be measured on a common momentum and energy grid by
\begin{align}
r_{\cal P}&=\frac{\|\Delta\mathcal A(\kk)+\Delta\mathcal A(-\kk)\|_2}
{2\|\Delta\mathcal A\|_2},\label{inversionResidualSM}\\
r_5&=\frac{\|\Delta\mathcal A(R_{2\pi/5}\kk)-\Delta\mathcal A(\kk)\|_2}
{2\|\Delta\mathcal A\|_2}.\label{fivefoldResidualSM}
\end{align}
Here \(\|\cdot\|_2\) is the discrete norm on that grid; absolute residuals are used when the denominator vanishes. {An exact finite-patch relation requires a symmetry of the full labeled Hamiltonian and of the boundary mask, not only of the uncolored vertex set. For the bulk construction proved in Sec.~\ref{sec:symmetry}, residuals can be monitored as the patch size or phason ensemble is increased at fixed broadening. Such convergence is not necessarily monotonic for individual cuts. These diagnostics test the numerical realization of the symmetry.} 
In addition, sample radius, boundary mask, broadening, harmonic cutoff, and momentum/energy resolution also control the finite-system result. 

\section{Symmetry origin of odd-parity spin splitting\label{sec:symmetry}}

Spin-group analyses of collinear odd-parity magnets emphasize that the orbital Hamiltonian must be included in the symmetry assessment, not only the ordered spin pattern \cite{Zeng2026SpinGroup, Ruanjiawei2026_PRL}. In particular, a spin-preserving antiunitary operation can be lost
through complex hopping even when the local moments remain collinear. Odd parity of the spin splitting does not assert physical time-reversal
invariance of that magnetic state. We apply this principle below to the full effective Hamiltonian, using quasicrystalline indistinguishability rather than importing a periodic spin space group.

\subsection{Indistinguishability and phase functions}

Because a quasicrystal is not invariant under a finite set of primitive
translations, its symmetry cannot be defined by exact pointwise
coincidence with a periodically repeated unit cell.  We write the spin
density in the spin-space-group framework \cite{Lifshitz1998,Lifshitz2004SpinGroups} as
\begin{equation}
 \bm S(\bm r)=\sum_{\kk\in\mathcal L}
 \bm S(\kk)e^{i\kk\cdot\bm r},                       \label{spinFourier}
\end{equation}
where the Fourier module \(\mathcal L\) contains all integral linear
combinations of the independent diffraction basis vectors.  Two spin
configurations are indistinguishable when they contain the same spatial
distribution of every bounded local pattern.  In Fourier space, each
symmetry or antisymmetry operation \([\gamma\Vert g]\) in the spin point
group \(G_S\) is then associated with a gauge phase function
\(\Phi_g^\gamma(\kk)\) such that
\begin{equation}
 \gamma\bm S(g\kk)=
 e^{2\pi i\Phi_g^\gamma(\kk)}\bm S(\kk),
 \qquad [\gamma\Vert g]\in G_S.                    \label{phasefunction}
\end{equation}
The phase allows the transformed structure to differ by a phason-space
translation while remaining physically indistinguishable.  It is the
appropriate replacement for the equality condition used in a periodic
crystal.

{For the bulk spectral constraints, rotations about the N\'eel axis form the spin-only subgroup \([SO(2)\Vert E]\). We call \(C_{2\perp}^{\rm s}\) a spin flip: it is a \(\pi\) spin rotation about an axis perpendicular to the N\'eel vector, exchanging the two conserved-spin sectors, rather than inversion of an arbitrary spin vector. The notation \(\mathcal P\) denotes the spatial operation constructed below. Their combination \([C_{2\perp}^{\rm s}\Vert\mathcal P]\) is termed the combined spin-flip--inversion relation throughout this section. In contrast, \([E_{\rm s}\Vert C_{5z}]\) preserves each spin sector. The spatial operations relate indistinguishable bulk configurations, not necessarily a finite sample to itself. Independently, the undriven real-hopping Hamiltonian has the spin-preserving antiunitary operation}
\[
\Theta_{\rm s}=C_{2\perp}^{\rm s}\TR
\]
{Taking \(C_{2\perp}^{\rm s}=-i\sigma_x\) and \(\TR=i\sigma_yK\), with \(K\) complex conjugation, gives \(\Theta_{\rm s}=i\sigma_zK\). It preserves the spin label, sends the plane-wave probe momentum to \(-\kk\), and satisfies \(\Theta_{\rm s}^2=+1\), so it does not itself enforce Kramers degeneracy. Combining its spectral relation with the bulk spin-flip--inversion relation gives the same-momentum opposite-spin spectral equality derived below.}

\subsection{Hyperspace action of inversion and fivefold rotation}

The sublattice label of a five-dimensional hyperlattice site is fixed by
the parity of its integer coordinates,
\begin{equation}
 {\bm n=\sum_{\nu=1}^{5}n_\nu\bm e_\nu}\in
 \begin{cases}
 A,&{\sum_\nu n_\nu}\in2\mathbb Z,\\
 B,&{\sum_\nu n_\nu}\in2\mathbb Z+1.
 \end{cases}                        \label{sublatticeParity}
\end{equation}

\subsubsection{An integer-lattice-compatible affine inversion}
The four occupied layers \(L_1,L_2,L_3,L_4\) have
\(q=\sum_i n_i=2,1,0,-1\), respectively. Write
\({S_\perp}\bm n=(\bm y,q/\sqrt2)\), with \(\bm y\)
the two transverse internal coordinates, and define their acceptance polygons by
\begin{equation}
W_q=\{\bm y:(\bm y,q/\sqrt2)\in W_\perp\},
\qquad q\in\{-1,0,1,2\}.
\label{magneticWindowsP}
\end{equation}
The window vertices in Sec.~\ref{sec:bipartite} occur in opposite pairs about
\(\bm c_\perp=(0,0,\sqrt2/4)\). Since the window is their convex hull,
\begin{equation}
2\bm c_\perp-W_\perp=W_\perp,
\qquad W_{1-q}=-W_q.
\label{windowCentralP}
\end{equation}
Thus the layer pairs \(L_1\leftrightarrow L_4\) and
\(L_2\leftrightarrow L_3\) have inverted acceptance polygons,
not merely equal total areas. The layers at the two axial tips of the
polytope have zero area and do not contribute to bulk frequencies.

Central symmetry of a window alone does not construct an operation on
the integer parent lattice. In particular, the linear map
\(\bm n\mapsto-\bm n\) preserves parity, whereas an affine map centered
at the window center need not preserve \(\mathbb Z^5\). An explicit
lattice-compatible choice is
\begin{align}
\mathcal P_\nu:\quad\bm n'&=\bm e_\nu-\bm n,\nonumber\\
\rr'&=\bm a_\nu-\rr,\qquad \bm a_\nu={S}\bm e_\nu,\nonumber\\
\rr_\perp'&={S_\perp}\bm e_\nu-\rr_\perp,
\qquad q'=1-q.
\label{affineParentP}
\end{align}
It is an involution on \(\mathbb Z^5\) and satisfies
\(\chi(\bm n')=-\chi(\bm n)\). Its physical action is inversion
about \(\bm a_\nu/2\), including the physical translation that accompanies
the parent-lattice shift. Different choices of \(\nu\) give equivalent
bulk relations; we suppress this choice in the symbol \(\mathcal P\).

Define \(\bm b_\nu\) by
\begin{equation}
(\bm b_\nu,0)={S_\perp}\bm e_\nu-2\bm c_\perp.
\label{internalShiftP}
\end{equation}
Its third component vanishes because the last row of
\({S_\perp}\) gives \(({S_\perp}\bm e_\nu)_3=1/\sqrt2\).
Equations~\eqref{windowCentralP}--\eqref{internalShiftP} imply
\begin{equation}
\bm y'=\bm b_\nu-\bm y,\qquad
W'_{1-q}=\bm b_\nu-W_q=W_{1-q}+\bm b_\nu.
\label{transformedWindowsP}
\end{equation}
Consequently, \(\mathcal P_\nu\) maps the selected tiling to one with
all four acceptance polygons translated by the same transverse internal
vector. It generally does not return the same vertex set pointwise.
The remaining step is to establish that this common translation
preserves the statistics of all labeled local patterns.

\subsubsection{Acceptance-domain proof of bulk indistinguishability}
We use the regular cut-and-project construction: the polygon boundaries
have zero area, and boundary choices do not affect bulk frequencies.
Within each layer, the parent points form a coset of
\(\{\bm d\in\mathbb Z^5:\sum_\nu d_\nu=0\}\).
Uniform distribution of their internal coordinates converts patch
frequencies into acceptance-domain areas, with a common density factor
for these cosets \cite{Moody2002Uniform}.

For a finite set \(\mathcal F\) of required occupied parent offsets,
including \(\bm d=0\), put \(q_{\bm d}=\sum_\nu d_\nu\) and
\(\bm d_\perp=({S_\perp}\bm d)_{1,2}\). The allowed internal
coordinates of an anchor in layer \(q\) form
\begin{equation}
\Omega_{q,\mathcal F}
=\bigcap_{\bm d\in\mathcal F}
 (W_{q+q_{\bm d}}-\bm d_\perp),
\label{patchDomainP}
\end{equation}
where an unoccupied layer has an empty window. Specifying an exact
local pattern additionally excludes offsets that must be absent;
these exclusions are represented by set differences of the same
translated polygons. For \(W_q^{(\bm v)}=W_q+\bm v\), both intersections
and differences obey
\begin{equation}
\Omega_{q,\mathcal F}^{(\bm v)}
=\Omega_{q,\mathcal F}+\bm v,\qquad
|\Omega_{q,\mathcal F}^{(\bm v)}|
=|\Omega_{q,\mathcal F}|.
\label{patchMeasureP}
\end{equation}
The same identity holds for domains including absent offsets.
More explicitly, the frequency per vertex of a specified magnetically
labeled pattern \(\mathfrak p\) is
\begin{equation}
\operatorname{freq}(\mathfrak p)
=\frac{\sum_{q\in Q_{\mathfrak p}}|\Omega_{q,\mathfrak p}|}
{\sum_{q=-1}^{2}|W_q|},
\label{normalizedPatchP}
\end{equation}
where \(Q_{\mathfrak p}\) contains the anchor layers compatible with
its magnetic labels and \(\Omega_{q,\mathfrak p}\) includes all required
occupancies and exclusions. The common coset density cancels between
numerator and denominator. Equation~\eqref{patchMeasureP} preserves
both, so a common transverse translation preserves the frequency of
every finite pattern, not only the populations of \(A\) and \(B\).

Under Eq.~\eqref{affineParentP}, an offset \(\bm d\) becomes
\(-\bm d\), and an anchor layer \(q\) becomes \(1-q\). The exchange
field at the image site is restored by the independent spin flip,
because \(\chi'=-\chi\) and
\(C_{2\perp}^{\rm s}\sigma_z(C_{2\perp}^{\rm s})^\dagger=-\sigma_z\).
Applying Eq.~\eqref{patchMeasureP} to Eq.~\eqref{transformedWindowsP}
therefore establishes indistinguishability of the complete labeled
nearest-neighbor magnetic network under the combined operation.
This conclusion uses the stated regular window and its parity coloring;
it is not inferred from the layer permutation alone, and it need not
hold for independently modified magnetic acceptance domains.

\subsubsection{Complex virtual hopping and the spectral consequence}
For a virtual path \(i-j-k\), affine inversion reverses both outward
bond vectors without reversing time or complex conjugating amplitudes.
Hence
\begin{equation}
\theta_{ij}'=\theta_{ij}+\pi,\quad
\theta_{kj}'=\theta_{kj}+\pi\pmod{2\pi},\qquad
\Lambda(\theta_{kj}'-\theta_{ij}')
=\Lambda(\theta_{kj}-\theta_{ij}).
\label{pathAmplitudeP}
\end{equation}
The coefficient \(i\eta\Lambda\) is therefore unchanged at fixed light
helicity. The intermediate vertex is mapped together with the endpoints,
so the acceptance domain contains the same transformed path constraints.
All paths contributing to an endpoint pair are added coherently;
their multiplicities and cancellations are carried into the image
network. Because the nearest-neighbor and retained virtual-hopping
amplitudes are deterministic local functions of this geometry, the
acceptance-domain result establishes equal bulk frequencies for the
full magnetically labeled hopping network of the order-\(1/\omega\)
Hamiltonian, including its complex weights.

Finally, \(\rr'=\bm a_\nu-\rr\) takes a plane-wave probe at \(\kk\)
to one at \(-\kk\), up to the common phase
\(e^{i\kk\cdot\bm a_\nu}\). This phase cancels from spectral weights.
The combined operation therefore gives
\(\overline{\mathcal A}_s(\kk,\varepsilon)
=\overline{\mathcal A}_{-s}(-\kk,\varepsilon)\), as derived from
spectral moments below. This is a bulk statistical relation for the
canonical regular construction, not a claim that the inversion center
\(\bm a_\nu/2\) is an exact symmetry center of every finite realization.

\subsubsection{Fivefold rotation}

The noncrystallographic fivefold rotation is a crystallographic
permutation in the five-dimensional parent lattice:
\begin{align}
 C_{5z}:\quad
 {(\bm e_1,\bm e_2,\bm e_3,\bm e_4,\bm e_5)}
 &\to{(\bm e_2,\bm e_3,\bm e_4,\bm e_5,\bm e_1)},\nonumber\\
 (n_1,n_2,n_3,n_4,n_5)
 &\to(n_5,n_1,n_2,n_3,n_4).                         \label{C5hyper}
\end{align}
It leaves \(\sum_i n_i\) invariant and does not exchange \(A\) and
\(B\).

\subsection{Undriven and driven bulk spectral constraints}

\subsubsection{Undriven bulk spectral equality}
{Let \(X\) denote a magnetically labeled tiling together with its hopping amplitudes, and \(gX\) its image under the combined spin-flip--inversion operation. Equations~\eqref{transformedWindowsP}--\eqref{pathAmplitudeP} establish that \(X\) and \(gX\) have the same frequencies of every finite labeled hopping pattern for the regular Penrose construction specified here. We now derive the electronic spectral consequence of that result. Write \(\overline{\mathcal A}_s\) for the thermodynamic vertex-averaged spectrum or the corresponding hull-averaged spectrum. For a different window or magnetic decoration, the labeled-network equivalence must be checked anew before applying the same argument.}

The connection between pattern statistics and this electronic observable can be seen from the plane-wave spectral moments. {Write \(H_s\) for the relevant single-particle spin block, either undriven or the retained driven \(H_{{\rm F},s}\).} For finite-range, bounded hopping they are
\begin{equation}
m_{p,s}(\kk)=\lim_{N_{\rm site}\to\infty}
\frac{1}{N_{\rm site}}\sum_{ij}
e^{i\kk\cdot(\rr_j-\rr_i)}(H_s^p)_{ij}.
\label{bulkmomentsSM}
\end{equation}
For each finite integer \(p\geq0\), a term is a finite hopping walk, including on-site exchange insertions. Its displacement, magnetic labels, and amplitude are determined by a bounded local pattern. Equal frequencies of all such patterns therefore give the transformed moment relation, and hence the corresponding spectral-measure relation. Equivalently, the resolvent can be reconstructed from these moments for bounded \(H_s\). This justifies the bulk constraint
\begin{equation}
\overline{\mathcal A}_s(\kk,\varepsilon)
=\overline{\mathcal A}_{-s}(-\kk,\varepsilon).
\label{bulkspinversionSM}
\end{equation}
The broadening \(\Gamma>0\) in the spectral function smooths the same spectral measure.

Without light, \(H_s\) is real symmetric. A real eigenbasis gives equal squared plane-wave overlaps at \(\kk\) and \(-\kk\), so
\begin{equation}
\mathcal A_s(\kk,\varepsilon)=\mathcal A_s(-\kk,\varepsilon).
\label{realblocksSM}
\end{equation}
Unlike Eq.~\eqref{bulkspinversionSM}, this relation holds for any finite undriven patch with real hoppings and a real spatial mask. Combining the two relations gives
\begin{equation}
\overline{\mathcal A}_\uparrow(\kk,\varepsilon)
=\overline{\mathcal A}_\downarrow(\kk,\varepsilon).
\label{undrivenequalitySM}
\end{equation}
This is equality of the opposite-spin bulk spectral measures at the same probe momentum, not a claim of exact Kramers pairing in every finite realization.

\subsubsection{Pointwise symmetry and an independent bipartite check}
A conventional real-space \(\PT\) symmetry requires a site permutation \(p\) and a physical inversion center \(\rr_0\) such that
\[
\rr_{p(i)}=2\rr_0-\rr_i,\qquad
\chi_{p(i)}=-\chi_i,\qquad
t_{p(i)p(j)}=t_{ij}
\]
for the real undriven hoppings, with the boundary also mapped to itself. Only under these additional conditions does \(\PT\) act within the same finite Hamiltonian. For a real involutory inversion permutation and spin-\(1/2\) time reversal, \((\PT)^2=-1\), yielding exact eigenstate pairing. The representation \(i\sigma_y\tau_xK\) has this property in the two-sublattice quasi-band model, but is not a construction of the required site permutation for a finite Penrose patch.

The particular nearest-neighbor model also admits an algebraic check that is independent of inversion. In the \(A/B\) basis,
\begin{equation}
H_s=\begin{pmatrix}
sN I_A&M\\M^\dagger&-sN I_B
\end{pmatrix},
\label{bipartiteblocksSM}
\end{equation}
where \(M\) is the intersublattice hopping matrix. Each nonzero singular value \(v_\mu\) of \(M\) produces a two-dimensional block
\begin{equation}
H_s^{(\mu)}=\begin{pmatrix}sN&v_\mu\\v_\mu&-sN\end{pmatrix},
\qquad E_{\mu,\pm}=\pm\sqrt{N^2+v_\mu^2}.
\label{singularpairSM}
\end{equation}
These paired energies are spin-independent. Null states confined to \(A\) have energy \(sN\), and those confined to \(B\) have energy \(-sN\); unequal sublattice populations can therefore leave unmatched finite-sample levels. Moreover, eigenvalue pairing does not imply equal plane-wave spectral weights. Equation~\eqref{undrivenequalitySM}, rather than energy pairing alone, is the statement relevant to the bulk spectral spin contrast.

\subsubsection{Driven bulk constraints}

{For the monochromatic vector potential of the main text, the polarization rotation is characterized by the dimensionless axial vector}
\begin{equation}
\bm\kappa_{\rm L}
=\frac{\bA(t)\times\dot{\bA}(t)}{\omega A_0^2}
=\eta\hat{\bm z}.
\label{polarizationaxisSM}
\end{equation}
Here the dot denotes a time derivative, and the in-plane vector potential is embedded in three dimensions. This vector specifies the sense of rotation about the fixed sample normal. 
Under spatial inversion, both \(\bA\) and \(\dot{\bA}\) change sign, so their cross product is unchanged. Under time reversal, \(\bA(t)\to-\bA(-t)\) and \(\dot{\bA}(t)\to\dot{\bA}(-t)\), so \(\bm\kappa_{\rm L}\) changes sign. The fivefold rotation \(C_{5z}\) preserves \(\bm\kappa_{\rm L}\). These are transformation properties of the polarization rotation in the spatially uniform orbital-coupling model. 

{The induced imaginary hopping has the same inversion-even, time-reversal-odd character. Equation~\eqref{pathAmplitudeP} verifies this inversion property for each virtual path, while the acceptance-domain proof includes the path's occurrence frequency and intermediate-site constraint. Complex conjugation instead reverses the helicity-dependent imaginary hopping, so Eq.~\eqref{realblocksSM} is no longer a constraint at fixed light helicity. The combined spin-flip--inversion relation \([C_{2\perp}^{\rm s}\Vert\mathcal P]\) and the spin-preserving fivefold relation remain bulk constraints for the specified regular construction. The former enforces}
\begin{equation}
 {\Delta\overline{\mathcal A}}(\kk,\varepsilon)=-{\Delta\overline{\mathcal A}}(-\kk,\varepsilon),
 \qquad
 E_{\lambda s}(\kk)=E_{\lambda,-s}(-\kk).           \label{oddrelation}
\end{equation}
{Here \({\Delta\overline{\mathcal A}}=\overline{\mathcal A}_\uparrow-\overline{\mathcal A}_\downarrow\) is a bulk momentum-resolved electronic}
spin signal. 
The loss of the real-Hamiltonian relation removes the additional step that gave Eq.~\eqref{undrivenequalitySM}. A nonzero spin contrast is therefore allowed at generic \(\kk\), while Eq.~\eqref{oddrelation} fixes its odd parity. The energies in that equation denote matched bulk spectral or quasi-band branches. 
\[
 \Delta_\lambda(-\kk)=-\Delta_\lambda(\kk),\qquad
 \Delta_\lambda=E_{\lambda,+}-E_{\lambda,-}.
\]

{As a separate consistency check, spatial inversion in the reduced quasi-band model exchanges \(A/B\) and sends \(\kk\to-\kk\); its internal sublattice factor is \(\tau_x\). The translation associated with its physical inversion center contributes only a common phase, which drops out of the Hamiltonian conjugation. Up to an overall spinor phase, the combined spin-flip--inversion factor is \(U_{\rm inv}=\sigma_x\tau_x\). Since \(B_5(-\kk)=B_5(\kk)\) and \(D_5(-\kk)=-D_5(\kk)\), it satisfies}
\begin{equation}
{U_{\rm inv}\mathcal H^{(0)}(\kk)U_{\rm inv}^\dagger
=\mathcal H^{(0)}(-\kk).}
\label{reducedInversionSM}
\end{equation}
{This reproduces Eq.~\eqref{oddrelation} in the projected theory. 
Spin--orbit coupling mixes the conserved-spin sectors and requires a separate material-specific symmetry analysis.
Odd parity of the spectrum does not establish physical \(\TR\) symmetry of a state with fixed helicity and N\'eel order \cite{Luo2026HiddenZeeman}.
For a finite patch that does not implement the relevant operations
exactly, these spectral constraints should be tested with size or
phason averaging rather than assumed for every boundary.

The spatial rotation \([E_{\rm s}\Vert C_{5z}]\) leaves the spin sector unchanged. Combining it with \([C_{2\perp}^{\rm s}\Vert\mathcal P]\) yields
\begin{align}
{\overline{\mathcal A}_s(R_{2\pi/5}\kk,\varepsilon)}&={\overline{\mathcal A}_s(\kk,\varepsilon)},\\
{\overline{\mathcal A}_s(R_{\pi/5}\kk,\varepsilon)}&={\overline{\mathcal A}_{-s}(\kk,\varepsilon)}.
\label{spinexchangeSM}
\end{align}
The second relation follows because \(R_{\pi/5}=-R_{-4\pi/5}\), with \(R_{-4\pi/5}\) a fivefold rotation. Thus the spin sum is unchanged by \(36^\circ\), while the spin difference reverses sign. This is the unitary spin-group operation \([C_{2\perp}^{\rm s}\Vert C_{10z}]\); it need not be identified with physical \(C_{5z}\TR\) or \(C_{10z}\TR\) symmetry of the driven Hamiltonian. {Expanding either the bulk spin contrast \(\Delta\overline{\mathcal A}\) or the matched-branch splitting \(\Delta_\lambda\) in \(e^{i\ell\theta}\),} fivefold rotation requires \(\ell\in5\mathbb Z\), and odd parity requires odd \(\ell\). The allowed indices are therefore \(\ell=\pm5,\pm15,\pm25,\ldots\). The leading permitted angular form is \(h\)-wave, while the relative weights at finite momentum are determined by the spectrum.

\section{Successive acceptance-window partitions and quasi-band theory}
\label{sec:acceptanceWin}

The acceptance window selects the physical sites through their
perpendicular-space coordinates. Partitioning this window according to
local patches is an established geometric construction: regions associated
with a given patch are called acceptance domains
\cite{Koivusalo2021}. We use this geometric information to build
progressively more detailed projected electronic Hamiltonians, in the
spirit of earlier quasicrystal effective-model approaches
\cite{wang2022effective}. Our approximation is specified by the partition
rule, the associated basis states, and the projected hopping matrix
elements, which we define below.

\subsection{Window partitions and their statistical weights}

Let \(W_0\) be an acceptance window. A hyperlattice displacement
\(\bm d\) has projections \(\rr={S}\bm d\) and
\(\rr_\perp={S_\perp}\bm d\). For a site with internal coordinate
\(\bm y\), the shell signature is
\begin{equation}
 {\mathcal Q_m(\bm y)}=\{\bm d\in\mathbb Z^{d_{\rm h}}:
 {\|\bm d\|_1=m},\ \bm y+{S_\perp}\bm d\in W_0\}.
 \label{shellsignature}
\end{equation}
To make a nested hierarchy precise, retain the tuple
{\((\mathcal Q_1,\ldots,\mathcal Q_l)\)} through order \(l\), or equivalently refine the
previous partition by the new shell boundaries. Sites with the same
retained signature belong to the same subwindow \(W_l^j\).
Disconnected regions may share a label if their signatures coincide.
This definition records membership at allowed offsets; it does not
require a connected hopping path to every offset.

{Here \(d_{\rm h}=5\) for Penrose and \(4\) for Ammann--Beenker geometry. The shell index \(m\) is distinct from the layer index \(q\) in Sec.~\ref{sec:symmetry}, and partition order \(l\) is distinct from the angular harmonic \(\ell\). In this general window construction, \(\bm y\) denotes the full internal coordinate, including its layer for Penrose; the earlier \(\bm y\in W_q\) described only its transverse components.}

Let \(\mu_j=\mu(W_l^j)\). {The measure \(\mu(W)\) is the appropriate geometric window measure and \(\mu_0=\mu(W_0)\); it is unrelated to the chemical potential \(\mu\) in Sec.~\ref{sec:trARPES}.} Uniform sampling of a regular cut-and-project
window gives
\begin{equation}
 p_{l,j}=\lim\frac{N_{l,j}}{N_{\rm site}}
 =\frac{\mu_j}{\mu_0},\qquad \sum_jp_{l,j}=1.
 \label{windowweight}
\end{equation}
For {Ammann--Beenker} tiling, \(\mu\) is area in the two-dimensional perpendicular plane.
For the layered Penrose construction it is the induced measure on the
occupied layers described above. Magnetic parities are kept as distinct
labels at every order: the zeroth-order Penrose magnetic theory has one
window for each of \(A\) and \(B\), even before local-environment refinement.
The geometric order \(l\) is independent of the order in \(1/\omega\)
of the Floquet expansion.

\subsection{Confined plane-wave states and the Gram matrix}

For an arbitrary occupied window \(W\), define
\begin{equation}
 |\kk,W,s\rangle=\frac{1}{\sqrt{N_W}}
 \sum_{\bm y_i\in W}e^{i\kk\cdot{\rr_i}}|i,s\rangle .
 \label{confinedstate}
\end{equation}
At the same \(\kk\), disjoint windows have exactly zero overlap. More
generally, their thermodynamic overlap is
\begin{equation}
 \langle\kk,W_1,s|\kk,W_2,s'\rangle
 =\delta_{ss'}\frac{\mu(W_1\cap W_2)}
 {\sqrt{\mu(W_1)\mu(W_2)}}.
 \label{samekoverlap}
\end{equation}
{For a disjoint window partition, this identity supplies an orthonormal finite-dimensional basis at each \(\kk\); overlapping windows instead require an overlap matrix.} It does not imply orthogonality between different momenta:
\begin{equation}
 {\mathsf G}_{(\kk,j),(\bm p,b)}
 =\frac{\sum_{i\in W_l^j\cap W_l^b}
 e^{i(\bm p-\kk)\cdot{\rr_i}}}
 {\sqrt{N_{l,j}N_{l,b}}}.
 \label{Gram}
\end{equation}
Bragg-related components can give nonzero off-diagonal Gram entries.
We use each fixed-\(\kk\) subspace as a projected quasi-band model.
If several momenta are combined, their Gram matrix must be retained
and the generalized eigenproblem solved on its non-null subspace.
Increasing \(l\) alone is not a proof of cross-momentum orthogonality
or completeness in the thermodynamic limit.

The Fourier module labels diffraction vectors. A finite-sample
plane-wave spectral function, in contrast, can be evaluated for any
continuous \(\kk\in\mathbb R^2\). A projected hyperlattice momentum
sampling is a computational choice rather than a restriction on the
photoelectron momentum. We use a finite decagonal region for Penrose
plots, called the first pseudo-Brillouin zone (1PBZ) in the source
figures. It does not tile reciprocal space as an exact primitive cell,
and spectral information outside it remains physical.

\subsection{Projected hopping and the zeroth-order Patterson limit}

Write a directed hopping as
\(H_{\rr}=t_{\rr}\sum_i|i+\rr\rangle\langle i|\), restricted to
occupied endpoints. A source coordinate \(\bm y\in W_l^b\) must obey
\(\bm y+\rr_\perp\in W_l^a\). With the phases in
Eq.~\eqref{confinedstate},
\begin{equation}
 {\mathcal H^{(l)}_{ab}(\kk)}
 =\sum_{\rr}t_{\rr}e^{-i\kk\cdot\rr}
 \frac{\mu[W_l^a\cap(W_l^b+\rr_\perp)]}
 {\sqrt{\mu_a\mu_b}} .
 \label{partitionH}
\end{equation}
The phase follows directly from
{\(e^{-i\kk\cdot(\rr_i+\rr)}e^{i\kk\cdot\rr_i}\)}.
Since \(t_{-\rr}=t_{\rr}^*\) and overlap measures are translation
invariant, {\(\mathcal H^{(l)}_{ba}=\mathcal H^{(l)*}_{ab}\)}.
The on-site magnetic term is \(sN\chi_a\delta_{ab}\).

In the Floquet problem a hopping amplitude can depend on the
intermediate vertex, not only on its endpoint displacement. The general
matrix element is therefore
\begin{equation}
 {\mathcal H^{(l)}_{ab,s}(\kk)}=sN\chi_a\delta_{ab}
 +\sum_{\rr}\frac{e^{-i\kk\cdot\rr}}{\sqrt{\mu_a\mu_b}}
 \int_{W_l^b\cap(W_l^a-\rr_\perp)}
 d\mu(\bm y)\,t_{\rr}(\bm y).
 \label{weightedpartitionH}
\end{equation}
Here \(t_{\rr}(\bm y)\) sums all allowed paths with that displacement.
For two successive steps \(\bm d_1,\bm d_2\), the integration domain
also requires
\(\bm y+{S_\perp}\bm d_1\in W_0\).
Their complex amplitudes are accumulated before diagonalization.
This preserves the rhombus cancellation automatically. Refining by
path signatures makes these amplitudes piecewise constant.

For a single nonmagnetic window, Eq.~\eqref{partitionH} reduces to
\begin{equation}
 {\mathcal H^{(0)}(\kk)}=\sum_{\rr}t_{\rr}e^{-i\kk\cdot\rr}
 {\Pi_W(\rr)},\qquad
 {\Pi_W(\rr)}=
 \frac{\mu[W_0\cap(W_0+\rr_\perp)]}{\mu_0}.
 \label{PattersonH}
\end{equation}
{Here \(\Pi_W\) denotes the normalized window-overlap (Patterson) weight. 
For a pair of partition regions, \(\mu_a=\mu(W_l^a)\) and \(\mu_b=\mu(W_l^b)\).} For two magnetic windows the normalization is
\(\sqrt{\mu_\alpha\mu_\beta}\). A conditional bond frequency
\(N_{\beta\alpha}(\rr)/N_\alpha\) equals the normalized matrix
weight only for equal sublattice populations; otherwise the factor
\(\sqrt{N_\alpha/N_\beta}\) must be included.

\begin{figure}[t]
 \centering
 \includegraphics[width=0.75\linewidth]{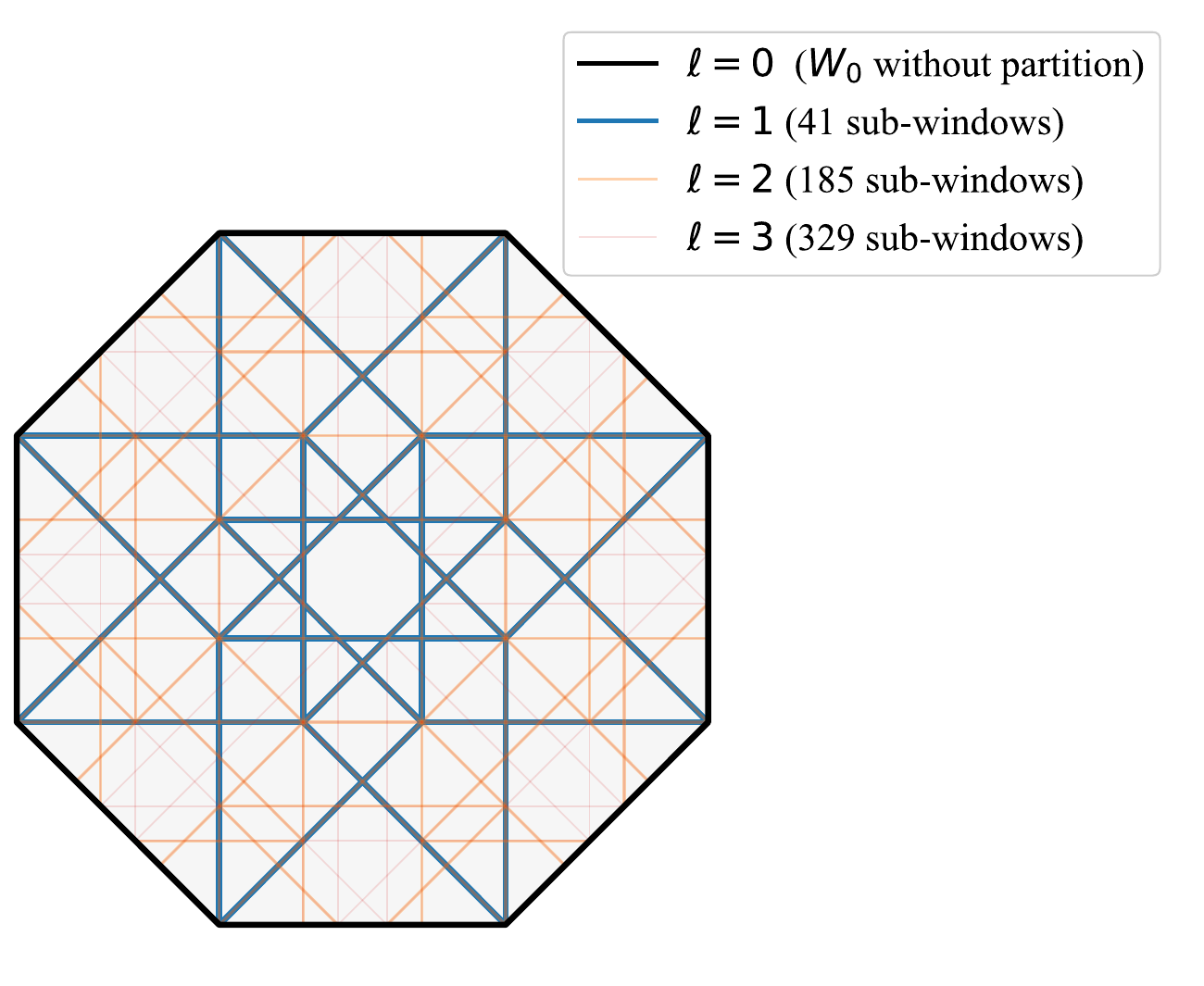}
 \caption{\label{fig:hierarchySM}Successive acceptance-window partitions
 in the AB benchmark. The diagram shows the unpartitioned
 octagon and the boundaries at orders \(1,2,3\), with reported region
 counts \(41,185,329\). These data illustrate the quasi-band method
 on an octagonal geometry.}
\end{figure}
\begin{figure}[t]
 \centering
 \includegraphics[width=0.8\linewidth]{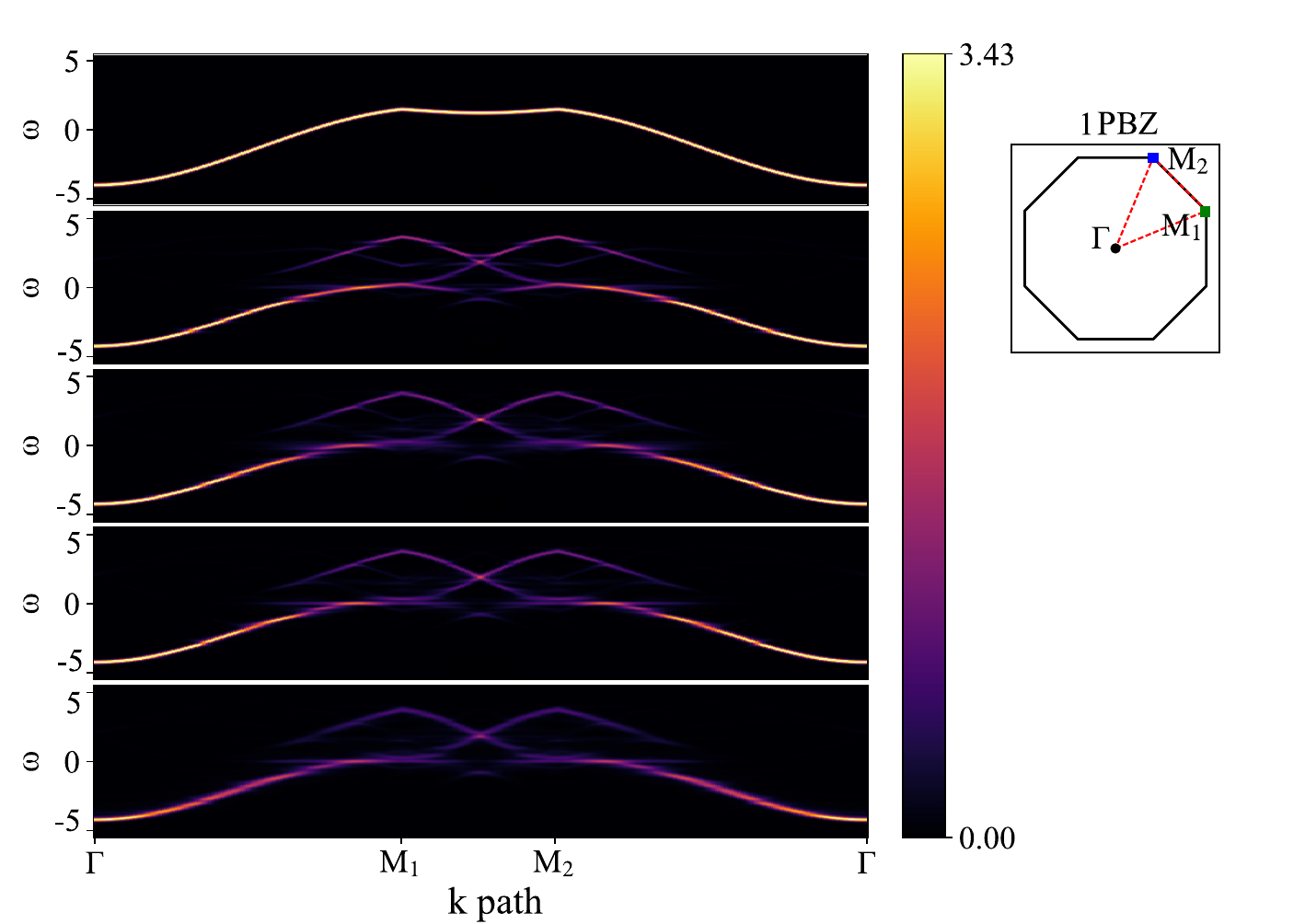}
 \caption{\label{fig:ABquasiband}AB spectral benchmark.
 From top to bottom: window orders \(0,1,2,3\), followed by exact
 diagonalization for \(N_{\rm site}=1433\). The path is indicated in
 the inset. The vertical symbol \(\omega\) in the original artwork
 denotes spectral energy in the benchmark units, not the CPL
 frequency of the Penrose model.}
\end{figure}

\begin{figure}[t]
 \centering
 \includegraphics[width=1\linewidth]{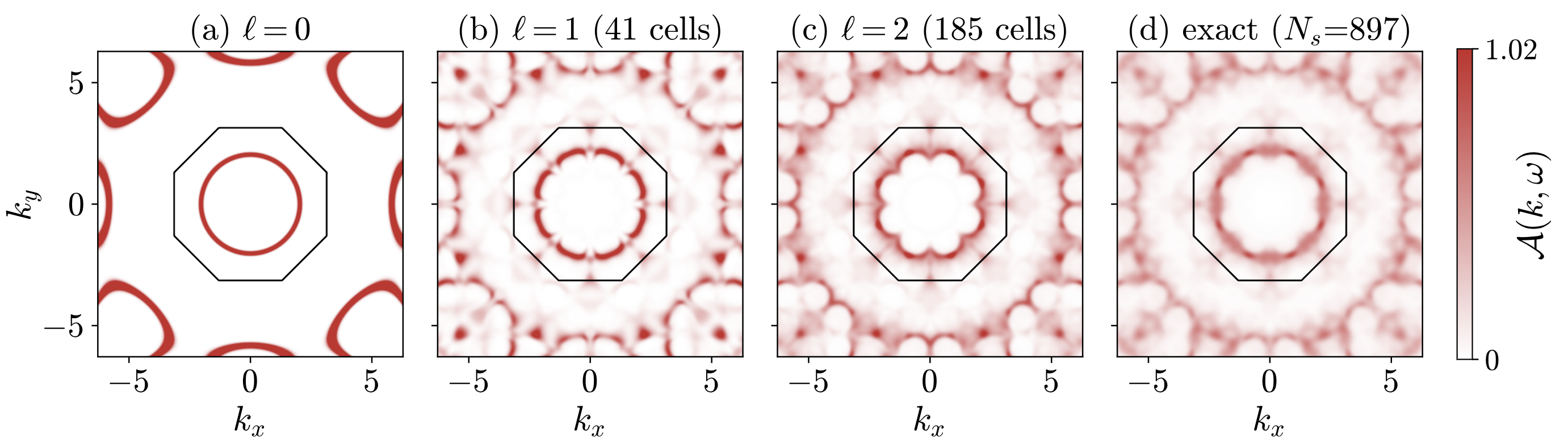}
 \caption{\label{fig:ABspectra}{AB spectral benchmark with spectral function distribution in 2D momentum plane.
 From left to right: window orders \(l=0,1,2\), followed by exact diagonalization for \(N_{\rm site}=897\).}}
\end{figure}

\begin{figure}[t]
 \centering
 \includegraphics[width=1.1\linewidth]{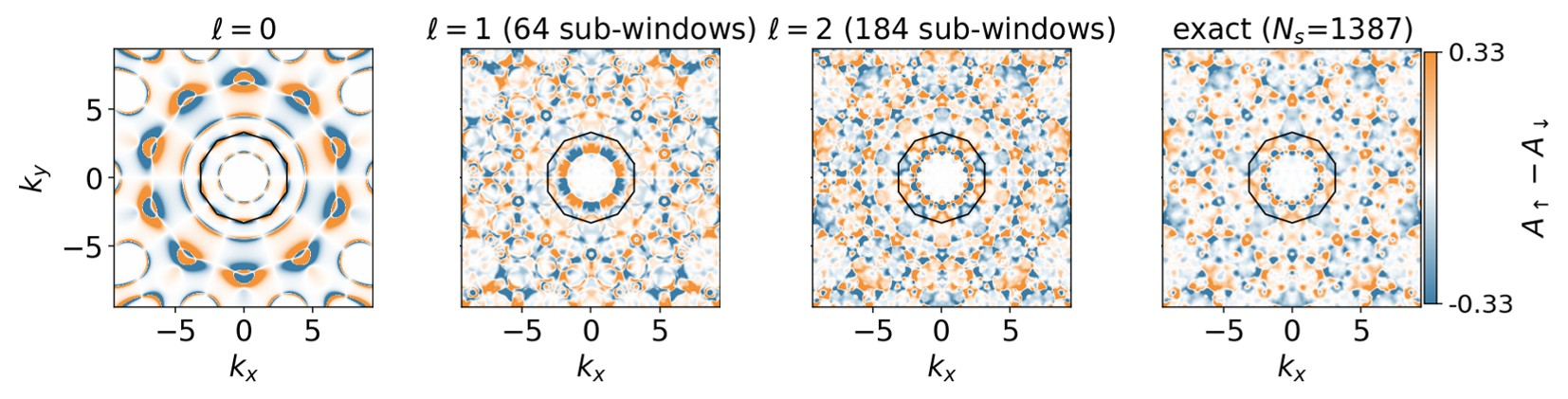}
 \caption{\label{fig:Penrosespectral}Penrose spectral benchmark with spin-difference spectral function distribution in 2D momentum plane.
 From left to right: window orders \(l=0,1,2\), followed by exact diagonalization for \(N_{\rm site}=1387\). }
\end{figure}

\subsection{Spectral weights and quasicrystal benchmarks}

If {\(\mathcal H^{(l)}(\kk)u_f=E_f u_f\)}, the unpartitioned plane wave satisfies
\begin{equation}
 |\kk,W_0\rangle=\sum_j\sqrt{p_{l,j}}\,|\kk,W_l^j\rangle.
 \label{coarsestate}
\end{equation}
Consequently, its projected spectral function is
\begin{equation}
 {\mathcal A^{(l)}(\kk,\varepsilon)}
 =-\frac1\pi\operatorname{Im}\sum_f
 \frac{\left|\sum_j\sqrt{p_{l,j}}\,u_{f,j}\right|^2}
 {\varepsilon-E_f+i\Gamma}.
 \label{hierarchyspectrum}
\end{equation}
There is one sum over bands \(f\): the window components combine
coherently inside the modulus. Completeness of the eigenvectors of the
projected matrix gives a total weight of one. For a spinful model the
same construction is performed in each conserved-spin block.

The {Ammann--Beenker} (AB) benchmark uses \(D_l=1,41,185,329\) {distinct subwindows} for orders \(l=0,1,2,3\), as shown in Fig.~\ref{fig:hierarchySM}. These partition sizes refer to AB geometry. Figure~\ref{fig:ABquasiband} compares their spectra to exact
diagonalization of a 1433-site AB cluster along \(\Gamma-M_1-M_2-\Gamma\). The zeroth order captures a broad dispersion; refinement produces additional branches and weight redistribution closer to the finite-system spectrum. Figure~\ref{fig:ABspectra} further shows the spectral function distribution at $\varepsilon=-0.8$ eV in the 2D momentum plane for orders $l=0,1,2$. In both figures, the refinement resolves additional local-environment features of the octagonal AB quasicrystal.

In addition, we examine the selection window partition hierarchy in Penrose quasicrystals. Figure~\ref{fig:Penrosespectral} compares the spin-difference spectra to exact diagonalization of a 1387-site Penrose quasicrystal cluster. With increasing order $l$, refinement produces additional peaks and weight redistribution closer to the exact cluster spectrum.

\subsection{Fivefold structure factors}

Let \(\bm a_j=a(\cos 2\pi j/5,\sin2\pi j/5)\), with indices understood
modulo five.  The nearest-neighbor part is summarized by
\begin{equation}
 C_5(\kk)=\frac{1}{5}\sum_{j=0}^4\cos(\kk\cdot\bm a_j),
 \qquad B_5(\kk)=4t_1J_0(\alpha_0)C_5(\kk).
 \label{B5SM}
\end{equation}
For a turning angle \(\delta\), the two-step displacement vectors form a
fivefold star of length \(L_\delta=2a\sin(\delta/2)\). We define
\[
 S_5(\kk;\delta)=\frac{1}{5}\sum_{j=0}^4
 \sin\!\left[
 2a\sin\!\left(\frac{\delta}{2}\right)
 \kk\cdot{\bm u_j}\right],
\]
where
\[
 {\bm u_j}=
 \left(
 \cos\!\left(\frac{(4j+1)\pi}{10}\right),
 \sin\!\left(\frac{(4j+1)\pi}{10}\right)
 \right)
\]
for the orientation convention of the main text.  Summing the
vertex-class contributions with Eq.~\eqref{vertexweights} gives
\begin{align}
 {D_5(\kk)}= &{(5\sqrt5-11)
 \sum_{\delta=2\pi/5,\,4\pi/5}\Lambda(\delta)S_5(\kk;\delta)},
 \label{D5full}\\
 I(\delta)=&{2\Lambda(\delta)}=\frac{4t_1^2}{\hbar\omega}
 \sum_{n=1}^{\infty}\frac{J_n^2(\alpha_0)}{n}\sin(n\delta).
 \label{Idelta}
\end{align}
One can check the geometric coefficient directly. With
\(X=I(2\pi/5)S_5(\kk;2\pi/5)\) and
\(Y=I(4\pi/5)S_5(\kk;4\pi/5)\), the contributions to the
\(B\)--\(B\) block at \(\eta=+1\) are
\begin{align}
 H_{BB}^{D,Q}&=0,\quad H_{BB}^{K}=2 \rho_KY,\nonumber\\
 H_{BB}^{S}&=5 \rho_SY,\quad
 H_{BB}^{S5}=-5 \rho_{S5}Y,\quad H_{BB}^{J}= \rho_JY,\nonumber\\
 H_{BB}^{T}&= \rho_T(X-5Y),\qquad
 H_{BB}^{V}= \rho_V(X-5Y).
 \label{vertexsum}
\end{align}
Equation~\eqref{vertexweights} gives
\( \rho_T+\ \rho_V=\xi\) and
\(2 \rho_K+5 \rho_S-5 \rho_{S5}
+ \rho_J-5 \rho_T-5 \rho_V=\xi\).
{Here \(\xi=(5\sqrt5-11)/2\), and \(I(\delta)=2\Lambda(\delta)\), with \(\Lambda\) defined by the virtual hopping Eq.~(3) in the main text.} Thus \(H_{BB}=\eta D_5\) and \(H_{AA}=-\eta D_5\). Here \(D_5\), {\(\Lambda\)}, and \(I\) are helicity independent; the only helicity
factor is the explicit \(\eta\). The resulting quasi-band Hamiltonian is
\begin{equation}
 {\mathcal H^{(0)}(\kk)}=
 N\sigma_z\tau_z
 -{B_5(\kk)}\tau_x
 -\eta D_5(\kk)\tau_z.                              \label{HeffSM}
\end{equation}
Its eigenvalues are
\[
 E_{\lambda s}(\kk)=\lambda
 {\sqrt{B_5^2(\kk)+[sN-\eta D_5(\kk)]^2}}.
\]
Here \(s=+1\) and \(-1\) correspond to \(\uparrow\) and \(\downarrow\), while \(\lambda=+1\) and \(-1\) label the upper and lower quasi-bands independently of spin. 

\section{Long-wavelength \textit{h}-wave expansion}
\label{sec:quasi-band}
The angular selection rule follows most transparently from complex
coordinates \(k_\pm=k_x\pm ik_y\).  Under a rotation by \(2\pi/5\),
\(k_\pm\to e^{\pm i2\pi/5}k_\pm\).  An odd scalar harmonic invariant
under fivefold rotations must have total odd degree and angular momentum
equal to a multiple of five.  Degrees one and three cannot satisfy this
condition.  At degree five the two real basis functions are
\begin{align}
 \operatorname{Re}k_+^5
 &=k_x^5-10k_x^3k_y^2+5k_xk_y^4
 =k^5\cos5\theta,                                   \label{cos5}\\
 \operatorname{Im}k_+^5
 &=5k_x^4k_y-10k_x^2k_y^3+k_y^5
 =k^5\sin5\theta.                                   \label{sin5}
\end{align}
Here \(\kk=k(\cos\theta,\sin\theta)\); \(\theta\) is the momentum
polar angle, not a directed real-space bond angle \(\theta_{ij}\).

The same result follows by expanding Eq.~\eqref{D5full}.  For a
fivefold star \(\bm u_j\),
\[
 \sum_j(\kk\cdot\bm u_j)^p=0
\]
for odd \(p=1,3\), while the \(p=5\) sum is proportional to
\(\cos5(\theta-\theta_0)\).  Hence
\begin{equation}
 D_5(\kk)=g_5(\alpha_0,\omega)
 k^5\sin(5\theta)+O(k^7).                 \label{D5small}
\end{equation}
{For the bond stars defined above, \(\theta_0=\pi/10\), so \(\cos[5(\theta-\theta_0)]=\sin(5\theta)\). This fifth harmonic changes sign under a \(36^\circ\) rotation.}
More explicitly, the elementary identity
\(\cos^5x=(10\cos x+5\cos3x+\cos5x)/16\) gives
\begin{equation}
 \frac15\sum_j\sin(L_\delta\kk\cdot\bm u_j)
 =\frac{L_\delta^5 k^5}{1920}
 \cos[5(\theta-\theta_0)]+O(k^7).
 \label{starTaylor}
\end{equation}
The positive fifth-order sign follows from the Taylor series of sine.
For the stars defined above, \(\theta_0=\pi/10\), and therefore
\begin{equation}
 {g_5=\frac{5\sqrt5-11}{1920}
 \sum_{\delta=2\pi/5,\,4\pi/5}\Lambda(\delta)L_\delta^5.}
 \label{g5explicit}
\end{equation}
For weak fields,
\(J_1(\alpha_0)\simeq\alpha_0/2\), so
\[
 g_5\propto \frac{t_1^2a^5\alpha_0^2}{\hbar\omega}.
\]
Symmetry permits this coefficient but does not require it to be nonzero
at every drive amplitude: Bessel-weighted contributions can cancel at
special parameters. Allowed angular indices are \(\ell=5,15,25,\ldots\);
radial corrections to \(\ell=5\) already enter at degree seven.
A \(36^\circ\) rotation maps a fivefold star onto its negative, giving
\(D_5(R_{\pi/5}\kk)=-D_5(\kk)\), while \(C_5\) is unchanged.
Thus \(E_{\lambda s}(R_{\pi/5}\kk)=E_{\lambda,-s}(\kk)\).

{Writing \(E_0(\kk)=\sqrt{N^2+B_5^2(\kk)}\), expansion of the matched-branch difference \(\Delta_\lambda=E_{\lambda,\uparrow}-E_{\lambda,\downarrow}\) for \(|D_5|\ll E_0\) gives}
\begin{equation}
 \Delta_\lambda(\kk)=
 -\frac{2\lambda\eta N D_5(\kk)}
 {E_0(\kk)}+O\!\left(D_5^3/E_0^2\right).               \label{splitting}
\end{equation}
This explicitly demonstrates the cooperative nature of the effect:
the nonrelativistic spin splitting vanishes if either \(N\) or the
light-induced \(D_5\) term is absent.

Since \(C_5(0)=1\) and \(C_5(\kk)=1+O(k^2)\), substituting
Eq.~\eqref{D5small} into Eq.~\eqref{splitting} gives, in the
\(\theta_0=\pi/10\) convention,
\begin{equation}
\Delta_\lambda(\kk)=
-\frac{2\lambda\eta N g_5}
{\sqrt{N^2+16t_1^2J_0^2(\alpha_0)}}
k^5\sin(5\theta)+O(k^7).
\label{finalsplitSM}
\end{equation}
Corrections to the even denominator enter at degree seven, whereas
the next odd power of \(D_5\) in the energy difference starts at
degree fifteen. This expression assumes an analytic, nondegenerate
spin-independent branch near the origin; the finite exchange
\(N\ne0\) ensures a nonzero denominator in the present model.

\section{Time-dependent pump and spin-resolved trARPES}
\label{sec:trARPES}

We use a finite-duration vector potential
\[
 \bA_{\rm pump}(t)=
 A_0{e^{-t^2/T_{\rm pump}^2}}
 [\cos\omega t,\eta\sin\omega t].
\]
{The pump maximum defines \(t=0\), as in the main text.} The probe envelope centered at delay \(t_d\) is
\[
 s_{\rm pr}(t,t_d)=e^{-(t-t_d)^2/T_{\rm probe}^2}.
\]
For independent electrons initially in thermal equilibrium, let
\({H_{\rm init}}|n\rangle={\epsilon_n}|n\rangle\), where \({H_{\rm init}}\) is
the single-particle Hamiltonian specifying the preparation at time
\(t_0\). The initial one-particle correlation matrix is
\begin{equation}
C_0=f({H_{\rm init}})=\sum_n f({\epsilon_n})|n\rangle\langle n|,
\qquad f(E)=[e^{(E-\mu)/(k_BT)}+1]^{-1}.
\label{initialoccupationSM}
\end{equation}
The matrix function applies the Fermi distribution to the eigenvalues,
not to individual matrix elements. In a site--spin basis,
\((C_0)_{ab}=\langle c_b^\dagger(t_0)c_a(t_0)\rangle\).
For \({H_{\rm init}}=V\,{\rm diag}({\epsilon_n})V^\dagger\), it is evaluated as
\(C_0=V\,{\rm diag}[f({\epsilon_n})]V^\dagger\). At zero temperature it is
the projector onto states below \(\mu\), with occupations at
\({\epsilon_n}=\mu\) specified by the chosen limiting prescription.
Unlike the many-body thermal density operator
\(\widehat\rho_0=Z^{-1}e^{-(\widehat H_{\rm init}-\mu\widehat N)/(k_BT)}\),
whose trace is one, \(C_0\) has trace equal to the mean particle number.

The single-particle propagator satisfies
\(i\hbar\partial_tU(t,t_0)={H(t)}U(t,t_0)\) and \(U(t_0,t_0)=I\),
where \({H(t)}\) includes the pump. The Heisenberg operators obey
\(\mathbf c(t)=U(t,t_0)\mathbf c(t_0)\). Substitution into
\(G^<_{ab}(t_1,t_2)=i\langle c_b^\dagger(t_2)c_a(t_1)\rangle\) gives
\begin{align}
G^<(t_1,t_2)&={iU(t_1,t_0)f(H_{\rm init})U^\dagger(t_2,t_0)},\nonumber\\
G^<_{ab}(t_1,t_2)&=i\sum_n f({\epsilon_n})
\psi_{n,a}(t_1)\psi_{n,b}^*(t_2),
\label{propagatedlesserSM}
\end{align}
{The first line is the main-text propagation formula, with \(C_0=f(H_{\rm init})\).} Here \(|\psi_n(t)\rangle=U(t,t_0)|n\rangle\). Each initial state
retains its statistical weight while its wavefunction evolves.
For a field-free equilibrium preparation, \({H_{\rm init}}\) is the
undriven Hamiltonian; it need not coincide with \({H(t_0)}\) if the
pump is already appreciable at the chosen starting time.
Here and in the propagation formulas, \(H(t)\), \(H_{\rm init}\), and the evolution operators act in single-particle space. The matrix \(H(t)\) is obtained from the coefficients of the quadratic Hamiltonian in the main text, with the finite pump substituted for the monochromatic field.

The spin-resolved photoemission intensity, neglecting matrix-element
variations, is
\begin{align}
 I_s(\kk,\varepsilon,t_d)
 \propto\operatorname{Im}\int dt_1dt_2\,
 &s_{\rm pr}(t_1,t_d)s_{\rm pr}(t_2,t_d)
 {e^{i\varepsilon(t_1-t_2)/\hbar}}\nonumber\\
 &\times G^<_{ss}(\kk;t_1,t_2).                     \label{trarpesSM}
\end{align}
An equivalent positive expression convenient for implementation is
\begin{equation}
 I_s(\kk,\varepsilon,t_d)\propto
 \sum_n f({\epsilon_n})\left|
 \int dt\,s_{\rm pr}(t,t_d)e^{i\varepsilon t/\hbar}
 \langle\kk,s|U(t,t_0)|n\rangle
 \right|^2 .
 \label{positiveARPES}
\end{equation}
This follows by inserting the occupied-state expansion of \(G^<\);
it supplies a positivity check on the double-time calculation.
{The time integral in Eq.~\eqref{positiveARPES} is a probe-filtered amplitude for each initially occupied state. Squaring it gives that state's photoemission contribution, and summing with the weights \(f({\epsilon_n})\) gives the total current. In the absence of the pump, \(U(t,t_0)|n\rangle=e^{-i{\epsilon_n}(t-t_0)/\hbar}|n\rangle\), so the current reduces to occupied plane-wave spectral weights broadened by the Fourier transform of the probe envelope. During the pump, the evolving wavefunctions carry both the changing electronic structure and the coherent time dependence. This is why the finite-pulse photocurrent need not coincide with the static van Vleck spectral function.}

The initial-state factors \(f({\epsilon_n})\) do not amount to assigning a
Fermi distribution to Floquet quasienergies. Occupations here are
propagated unitarily from the specified initial density matrix.
By contrast, Li, Shao, and Kovalev determine driven steady-state spin
responses with explicit bath coupling \cite{Li2026FloquetSpin}.
Their spin-conserving bosonic-bath and fermionic-reservoir settings
give different spin-generation and transport behavior. Consequently,
a spin-polarized photoemission pattern in our closed-system protocol
does not establish a steady net spin accumulation or spin current.
Such claims would require a specified relaxation model and a separate
calculation of the corresponding observable.

A midpoint propagation step is
\[
U(t_{j+1},t_j)\simeq
\exp[-iH((t_{j+1}+t_j)/2)\delta t/\hbar].
\]
{Refinement of \(\delta t=t_{j+1}-t_j\) controls the time-discretization error; \(U^\dagger U=I\) provides an independent unitarity check. Equation~\eqref{positiveARPES} evaluates the photocurrent through probe-filtered amplitudes of initially occupied states and is manifestly nonnegative.}

\subsection{Parameters and initialization}

The main-text photoemission map uses
\begin{center}
\begin{tabular}{ll}
\hline
Quantity & Value\\
\hline
\(t_1,\ N\) & \(1\,{\rm eV},\ 1\,{\rm eV}\)\\
\(\alpha_0\) & \(1.5\)\\
\(\omega\) & \(20\,{\rm Prad/s}=20\,{\rm rad/fs}\)\\
\(\hbar\) & \(0.6582\,{\rm eV\,fs}\)\\
\(\hbar\omega,\ 2\pi/\omega\) & \(13.164\,{\rm eV},\ 0.31416\,{\rm fs}\)\\
\(T_{\rm pump},\ T_{\rm probe}\) & \(150\,{\rm fs},\ 10\,{\rm fs}\)\\
{\(t_d\)} & {\(0\)}\\
\(T,\ \mu\) & \(0,\ 0\)\\
Energy of the momentum map & \(-1.20\,{\rm eV}\)\\
\hline
\end{tabular}
\end{center}
The probe is centered at the pump maximum, {\(t_d=0\)}. Initial thermal occupations are set by \(H_{\rm init}\) and are subsequently propagated, rather than rethermalized at each time. A field-free preparation and an equilibrium preparation in a finite initial vector potential are physically different protocols. The optical period \(2\pi/\omega=0.31416\,{\rm fs}\) sets the rapid timescale of the propagation.

\section{Drive regime and limitations}
\label{sec:limitation}

The van Vleck Hamiltonian is retained through order \(1/\omega\),
with corrections of order \(\omega^{-2}\) omitted
\cite{Eckardt2015}. Its high-frequency regime requires
\begin{equation}
\frac{t_1}{\hbar\omega}\ll1,\qquad
\frac{N}{\hbar\omega}\ll1,
\label{highfrequencySM}
\end{equation}
together with an off-resonant separation of the photon energy from
the electronic spectrum relevant to the driven model. The local
ratios in Eq.~\eqref{highfrequencySM} are useful scale estimates,
but do not by themselves bound the truncation error: coordination
and the full bandwidth also enter. For the photoemission parameters,
\(\hbar\omega=13.164\,{\rm eV}\) and both ratios are approximately
\(0.0760\). Quantitative accuracy of the effective spectrum is assessed
by comparison with higher-order terms or a converged Floquet calculation,
rather than by these ratios alone. No weak-field expansion in
\(\alpha_0\) is required when the Bessel sums are retained. At fixed
\(\alpha_0\), increasing \(\omega\) improves the high-frequency hierarchy
but reduces the leading virtual-hopping scale \(t_1^2/(\hbar\omega)\),
and hence the spin splitting.

The pulsed trARPES calculation instead propagates the time-dependent
Hamiltonian and includes micromotion within the specified independent-electron
model; it does not replace the pulse by a static \(H_{\rm F}\).
For the chosen envelopes, \(\omega T_{\rm pump}=3000\) and
\(\omega T_{\rm probe}=200\), so both envelopes vary slowly over an
optical cycle. This separation supports comparison with the effective
spectrum near the pump maximum, but does not imply adiabatic electronic
occupations or a thermal Floquet steady state. Time-step and initial-time
convergence are separate from convergence of the high-frequency expansion.

The probe duration also sets an energy-resolution scale. For the
Gaussian amplitude envelope used here, an isolated stationary level has
the probe-limited intensity full width at half maximum
\begin{equation}
\delta\varepsilon_{\rm pr}
=\frac{2\sqrt{2\ln2}\,\hbar}{T_{\rm probe}}
\simeq0.155\,{\rm eV}
\quad (T_{\rm probe}=10\,{\rm fs}).
\label{proberesolutionSM}
\end{equation}
This follows from the squared Fourier transform of the probe envelope.
Pump-induced evolution, scattering, and instrumental resolution can
further modify the line shape. A nonzero spin-difference intensity
therefore need not imply two separately resolved opposite-spin peaks.
The probe width is distinct from the Lorentzian broadening \(\Gamma\)
used in the static spectral function.

Converting the dimensionless drive to a physical field requires the
edge length \(a\). The electric-field amplitude at the pump center is
\(\mathcal E_0=\omega A_0\), distinct from the quasi-band energy scale
\(E_0\), and satisfies
\begin{equation}
\mathcal E_0a=\frac{\alpha_0\hbar\omega}{e}
=19.746\,{\rm V}\qquad(\alpha_0=1.5).
\label{fieldscaleSM}
\end{equation}
The quoted parameters specify a model calculation, not an established
experimental operating window. In a material, optical absorption,
transitions outside the retained electronic subspace, and the stability
of the magnetic moments must also be considered. Being off-resonant
within the model does not establish negligible heating in the material.

The present treatment assumes conserved spin, independent electrons
coupled to a fixed staggered exchange field, and no bath or
self-consistent magnetic dynamics. It therefore addresses transient
electronic spin splitting and photoemission contrast, rather than the
formation or long-time stability of magnetic order. The coarsest
quasi-band model additionally averages local environments, whereas the
finite real-space calculations retain their individual hopping paths
and diffraction structure. Refining the window partition addresses the
former approximation; increasing the patch size and controlling its
boundary address the latter finite-size effects. The bulk symmetry
relations constrain the allowed angular harmonics as long as the
assumed spin-group relations hold, but do not determine their relative
weights, linewidths, or experimental visibility. Spin--orbit coupling,
interaction-induced relaxation, and changes of magnetic order require
a corresponding extension of the model and its symmetry analysis.

\end{document}